\documentclass[12pt,aps,nofootinbib,superscriptaddress]{revtex4}

\usepackage{amsmath,amssymb,bm,hyperref}
\usepackage[dvips]{graphicx}
\usepackage{hyperref}
\usepackage{yfonts}
\usepackage{color}
\newcommand{\beq}{\begin{eqnarray}}
\newcommand{\eeq}{\end{eqnarray}}
\renewcommand\d{\partial}

\newcommand{\tr}{\mathop{\mathrm{tr}}}
\newcommand{\SU}{\text{SU}}
\newcommand{\U}{\text{U}}

\begin{document}

\title{Chiral Langevin theory for non-Abelian plasmas}

\author{Yukinao Akamatsu}
\affiliation{Kobayashi-Maskawa Institute for the Origin of Particles and the Universe,
Nagoya University, Nagoya 464-8602, Japan}

\author{Naoki Yamamoto}
\affiliation{Maryland Center for Fundamental Physics, 
Department of Physics, University of Maryland,
College Park, Maryland 20742-4111, USA}

\begin{abstract}
Charged plasmas with chirality imbalance are unstable and tend to reduce the imbalance. 
This chiral plasma instability is, however, not captured in (anomalous) hydrodynamics 
for high-temperature non-Abelian plasmas. We derive a Langevin-type classical effective 
theory with anomalous parity-violating effects for non-Abelian plasmas that describes the 
chiral plasma instability at the magnetic scale. We show that the time scale of the instability 
is of order $[g^4 T \ln(1/g)]^{-1}$ at weak coupling.
\end{abstract}

\maketitle

\section{Introduction}
\label{sec:introduction}
Matter with chirality imbalance is expected to appear in a wide range of physical systems, 
such as quark-gluon plasma created in relativistic heavy ion collision experiments 
at the Relativistic Heavy Ion Collider (RHIC) and the Large Hadron Collider (LHC)
\cite{Kharzeev:2007tn, Kharzeev:2007jp}, the electroweak plasma in 
the early Universe \cite{Joyce:1997uy}, and electron plasmas inside neutron 
stars \cite{Charbonneau:2009ax}, among others. 
One of the direct consequences of the chirality imbalance is an unusual transport 
phenomena in the presence of a strong magnetic field ${\bm B}$ (which is also 
expected in most of these systems) called the chiral magnetic effect 
\cite{Vilenkin:1980fu,Nielsen:1983rb,Alekseev:1998ds,Son:2004tq,Fukushima:2008xe}: 
\beq
\label{CME}
{\bm j} = \frac{e^2\mu_5}{2\pi^2} {\bm B} \equiv \kappa {\bm B}.
\eeq
Here ${\bm j}$ is the electric current and $\mu_5$ is the chiral chemical potential
that characterizes the chirality imbalance. This current is unusual in that it flows in the 
direction of the magnetic field and it is dissipationless. Both facts are indeed related to 
the quantum anomalies \cite{Adler,BellJackiw}, which are an intrinsic property of 
relativistic quantum field theories. The chiral magnetic effect may have been observed 
in heavy ion collision experiments at RHIC%
\footnote{
For the current status, see Refs.~\cite{Kharzeev:2013ffa, Liao:2014ava} and references therein.
}
and may be observable in new materials called Weyl semimetals \cite{Vishwanath, BurkovBalents, Xu-chern}.

Recently, it was argued based on the kinetic theory with Berry curvature corrections
(or simply chiral kinetic theory) \cite{Son:2012wh, Stephanov:2012ki, Son:2012zy, Chen:2012ca}
that charged plasmas with chirality imbalance have unstable infrared collective modes 
that tend to reduce the imbalance, and thus, quark-gluon plasmas exhibiting the 
chiral magnetic effect are not stable \cite{Akamatsu:2013pjd}. 
This is the chiral plasma instability.%
\footnote{
One might wonder why we call it the \emph{plasma instability}. 
Our terminology parallels the conventional Weibel plasma instability \cite{Weibel} which 
occurs when the distribution function of particles is anisotropic in momentum space. This is 
manifested in tachyonic collective modes of gauge fields that tend to make the distribution 
function isotropic. Similarly, the chiral plasma instability appears when fermions have chiral 
asymmetry, as is manifested in tachyonic modes that tend to make the numbers of right- 
and left-handed fermions equal \cite{Akamatsu:2013pjd}. 
In other words, the chiral plasma instability can be regarded as a generalization of the conventional 
plasma instability to the case where the distribution function of fermions and the polarization 
tensor of gauge fields are parity noninvariant \cite{Akamatsu:2013pjd}.
}
A related instability has also been pointed out for the electroweak theory at high density 
\cite{Redlich:1984md, Tsokos:1985ny, Niemi:1985ir, Rubakov:1985nk, Rubakov:1986am}
and for the primordial magnetic field in the early Universe \cite{Joyce:1997uy, Boyarsky:2012ex}, 
using different theoretical frameworks. Notably, the instability was analyzed based on the 
\emph{nonlocal} hard dense loop effective action in Ref.~\cite{Laine:2005bt}. 
(For recent works on other theoretical issues, see 
Refs.~\cite{Khaidukov:2013sja, Jensen:2013vta, Avdoshkin:2014gpa}.)
In this paper, we shall concentrate on high-temperature non-Abelian plasmas with 
chirality imbalance, whose dynamics will turn out to be richer than Abelian ones.

It would be useful to have a simple local effective theory (simpler than the kinetic theory) 
to describe the dynamical evolution of non-Abelian chiral plasmas. 
To this end, one might come up with hydrodynamics, which is an effective theory valid at 
long distances and long time scales compared with the mean free path, $l_{\rm mfp}$, 
and mean free time, $\tau_{\rm mfp}$. Indeed, it has proven quite useful for describing 
a quark-gluon plasma after thermalization. In the hydrodynamic description for 
high-temperature non-Abelian plasmas, we need not care about non-Abelian gauge fields, 
because both chromoelectric and chromomagnetic fields are screened at scales 
(inverse of the Debye mass $\sim gT$ and the magnetic scale $\sim g^2 T$, respectively) 
much shorter than the mean free path, $l_{\rm mfp} \sim (g^4 T)^{-1}$, 
as we shall review later in Sec.~\ref{sec:scales}.
This is the reason why the effective theory must be hydrodynamics rather than 
\emph{chromomagneto}-hydrodynamics.%
\footnote{
In contrast, Abelian plasmas do not have the magnetic screening, and magnetic 
fields persist even in the hydrodynamic regime; this is why the long-range effective theory for 
Abelian plasmas is \emph{magneto}-hydrodynamics, in which case the Abelian chiral plasma 
instability can be captured.
} 
However, as was found in Ref.~\cite{Akamatsu:2013pjd} (see also below), 
it is exactly this magnetic scale where the non-Abelian chiral plasma instability occurs 
in the presence of the chirality imbalance, $\mu_5 \sim T$.%
\footnote{
Here and below we mean by ``$\sim$"  that they are parametrically of the same order in $g$.
}
This means that the dynamics of non-Abelian chiral plasma instability is completely 
missed in the hydrodynamic description. This is so even in the anomalous hydrodynamics \cite{Son:2009tf} which was recently constructed after the finding in the gauge/gravity duality 
computations of Refs.~\cite{Erdmenger:2008rm, Banerjee:2008th};
see also Refs.~\cite{Banerjee:2012iz, Jensen:2012jy, Haehl:2013hoa} for other formulations 
and Ref.~\cite{Hongo:2013cqa} for a numerical analysis of anomalous hydrodynamics.

The purpose of this paper is to construct an effective theory at the magnetic scale
that describes the dynamical evolution of non-Abelian chiral plasmas. 
Our main result is given by the Langevin-type equation for the gauge field, 
Eq.~(\ref{chiral_Langevin}) [which can be equivalently rewritten in the form of 
Eqs.~(\ref{Langevin}) and (\ref{CS})], supplemented by the evolution equation for the 
chiral charge, Eq.~(\ref{eq:n_5}). This is a generalization of the Langevin-type equation
without anomalous effects previously derived in 
Refs.~\cite{Bodeker:1998hm, Arnold:1998cy, Litim:1999ns}.
When we are only interested in the physics at the magnetic scale, our effective theory is 
much simpler than the chiral kinetic theory that also includes the (semi)hard degrees of 
freedom with momenta $k \gtrsim gT$.
It is also cheaper in performing practical numerical simulations than the chiral kinetic theory, 
since the former depends only on the coordinate ${\bm x}$, while the latter depends on both 
${\bm x}$ and momentum ${\bm p}$.
On the basis of this Langevin theory, we shall give parametric estimates of the time scale of 
the chiral plasma instability and that of the variation of the chiral charge.%
\footnote{
In a quark-gluon plasma with chirality imbalance, both photons and gluons 
that mediate the electromagnetic and strong forces become tachyonic,
because quarks have both $\U(1)$ electric and $\SU(3)$ color charges; there are Abelian and 
non-Abelian chiral plasma instabilities at the same time in this case. Here we are interested 
in the non-Abelian chiral plasma instability that is not captured by hydrodynamics.
}

Throughout this paper, we assume massless fermions, which should be a good approximation
when the temperature $T$ is sufficiently large compared with the fermion mass $m$.
We also assume $\mu_5 \sim T$.

\subsection{Hierarchy of scales and effective theories}
\label{sec:scales}
We pause here to recapitulate the hierarchy of scales in high-temperature non-Abelian 
plasmas and their low-energy effective theories. 
(For a pedagogical review, see Ref.~\cite{Arnold:2007pg}.)
We consider a sufficiently high-temperature regime so that the coupling constant is small, 
$g \ll 1$, where there is a definite separation of momentum scales: 
$g^4 T \ll g^2 T \ll gT \ll T$ (up to logarithmic corrections). In the following, we will refer 
to the scales $T$, $gT$, and $g^2 T$, as the hard, semihard, and soft (or magnetic) 
scales, respectively. Among others, it is easy to see that that the gauge boson acquires a 
semihard Debye screening mass $\sim gT$ in medium by perturbatively computing one-loop 
Feynman diagrams. However, the magnetic scale $g^2 T$ emerges nonperturbatively 
and it cannot be understood in an analogous way.

To understand the origin of the magnetic scale, consider the regime where the amplitude 
of fluctuations of the gauge field $A(k)$ with momentum $k$ becomes nonperturbative. 
This amounts to the condition that the contributions of nonlinear interactions are comparable 
to those of linear ones, $kA \sim g A^2$. On the other hand, the law of equipartition of energy 
states that the magnetic field has the energy $\sim T$, and hence, $B^2 R^3 \sim T$, where
$R \sim 1/k$ is the typical spatial size. Combining these two conditions, the typical scales 
of the momentum and the gauge field read
\beq
\label{scale}
k \sim g^2 T, \qquad A \sim gT, \qquad B \sim g^3 T^2.
\eeq
In this way, the magnetic (or the soft) scale $k \sim g^2 T$ emerges.

Let us then explain that the mean free path is given by $l_{\rm mfp} \sim (g^4 T)^{-1}$
(up to logarithmic corrections). Remember first the textbook formula, $l_{\rm mfp} \sim (n \sigma)^{-1}$,
where $n \sim T^3$ is the density of scatterers and $\sigma$ is the cross section. 
Because the scattering amplitude involves $g^2$, the cross section must be proportional 
to $g^4$. Note also that the cross section is proportional to $l^2$, where the length scale 
$l$ is given by the inverse of the typical momentum exchange, $l \sim T^{-1}$. Hence, 
we have $\sigma \sim g^4 l^2 \sim g^4/T^2$. Inserting them into the formula above, 
we get $l_{\rm mfp} \sim (g^4 T)^{-1}$.

\begin{figure}[t]
\begin{center}
\includegraphics[width=9cm]{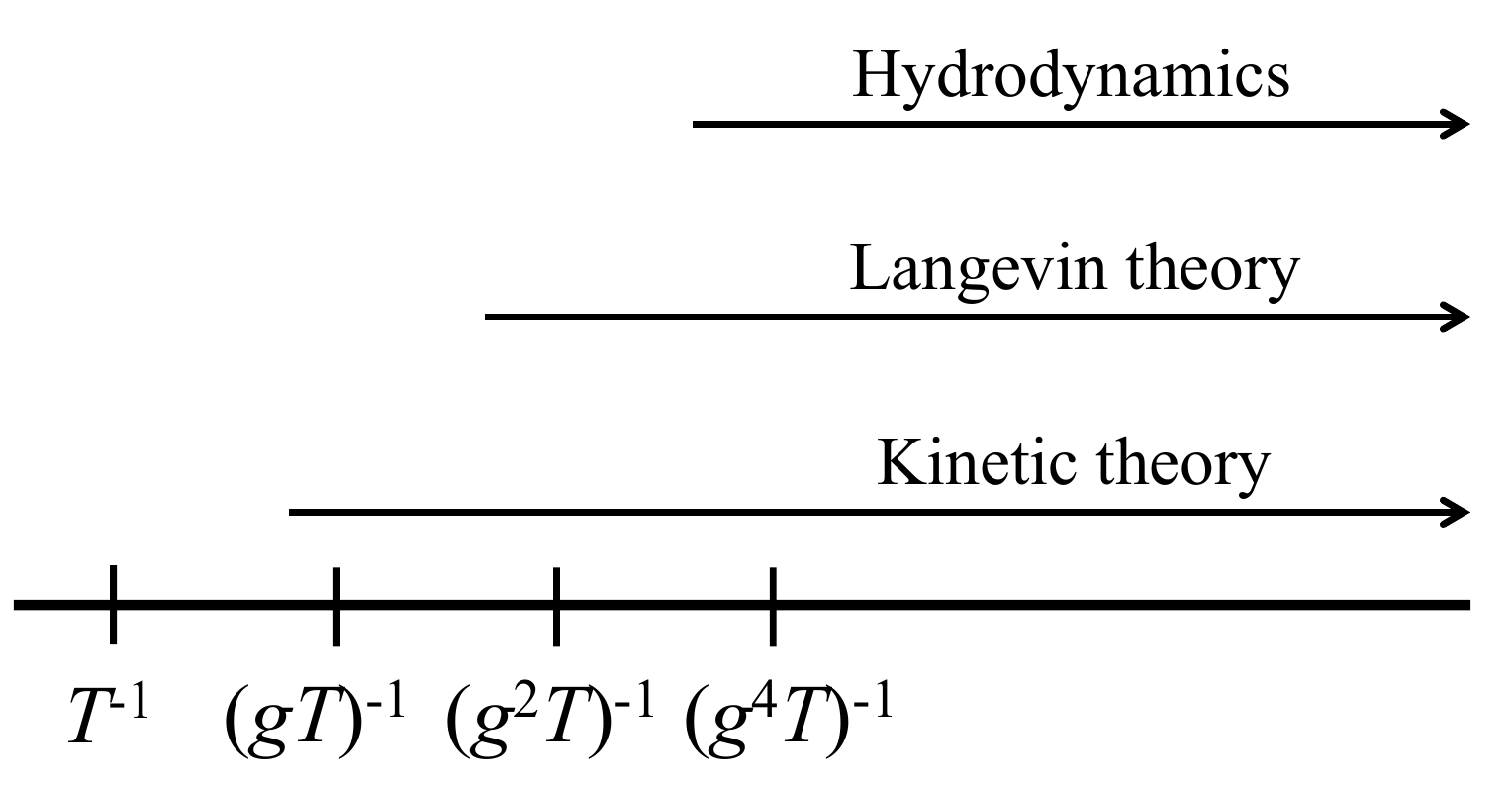}
\end{center}
\vspace{-0.5cm}
\caption{Hierarchy of length scales and applicability of effective theories in
high-temperature non-Abelian plasmas at weak coupling.} 
\label{fig:scales}
\end{figure}

In summary, we have a hierarchy of length scales as depicted in Fig.~\ref{fig:scales}.
Due to the screening effects, there are no chromoelectric and chromomagnetic fields at 
$L \gg (gT)^{-1}$ and $L \gg (g^2 T)^{-1}$, respectively. Kinetic theory describes
the dynamics at $L \gtrsim (gT)^{-1}$, while hydrodynamics is only applicable at 
$L \gg l_{\rm mfp} \sim (g^4 T)^{-1}$. The Langevin theory that we shall derive in this paper 
describes the physics at $L \gtrsim (g^2 T)^{-1}$, which has some overlapping regime with 
the kinetic theory, but is beyond the applicability of hydrodynamics. It is this scale where 
the non-Abelian chiral plasma instability occurs, as we are going to explain in a moment below.

\subsection{Intuitive picture of the chiral plasma instability}
\label{sec:physics}
We now give an intuitive argument of the chiral plasma instabilities.%
\footnote{
The presence of the instability for an Abelian chiral plasma was shown in 
Ref.~\cite{Joyce:1997uy} using the anomalous Maxwell equations 
(Maxwell equations plus the chiral magnetic effect). 
However, we are not aware of any previous, physical argument, 
and we believe it worthwhile to provide it here explicitly.
} 
Here we consider an Abelian plasma for simplicity, but our argument 
is also applicable to non-Abelian plasmas as long as the amplitude of the gauge fields is 
sufficiently small such that the nonlinear gauge interactions are negligible 
(in which case the Yang-Mills equation reduces to the Maxwell equation).
This argument is based only on the basic laws of electromagnetism (Maxwell equations) 
and the chiral magnetic effect in Eq.~(\ref{CME}), assuming a homogeneously distributed 
chirality imbalance, $\mu_5 > 0$, in the initial state.
For further simplicity, we ignore the effects of dissipation and do not consider the 
Ohmic current ${\bm j}_{\rm ohm} = \sigma {\bm E}$. These effects will be discussed
in detail in the following sections.%
\footnote{
For non-Abelian plasmas, the effects of dissipation are essential at the quantitative 
level: the typical frequency of the chiral plasma instability would be $\omega \sim g^4 T$ 
without dissipation, while it is $\omega \sim g^4 T \ln(1/g)$ with dissipation 
\cite{Akamatsu:2013pjd}; see also below. The argument here provides, at least, 
a qualitative understanding of the emergence of the chiral plasma instability.
}

\begin{figure}[t]
\begin{center}
\includegraphics[width=9cm]{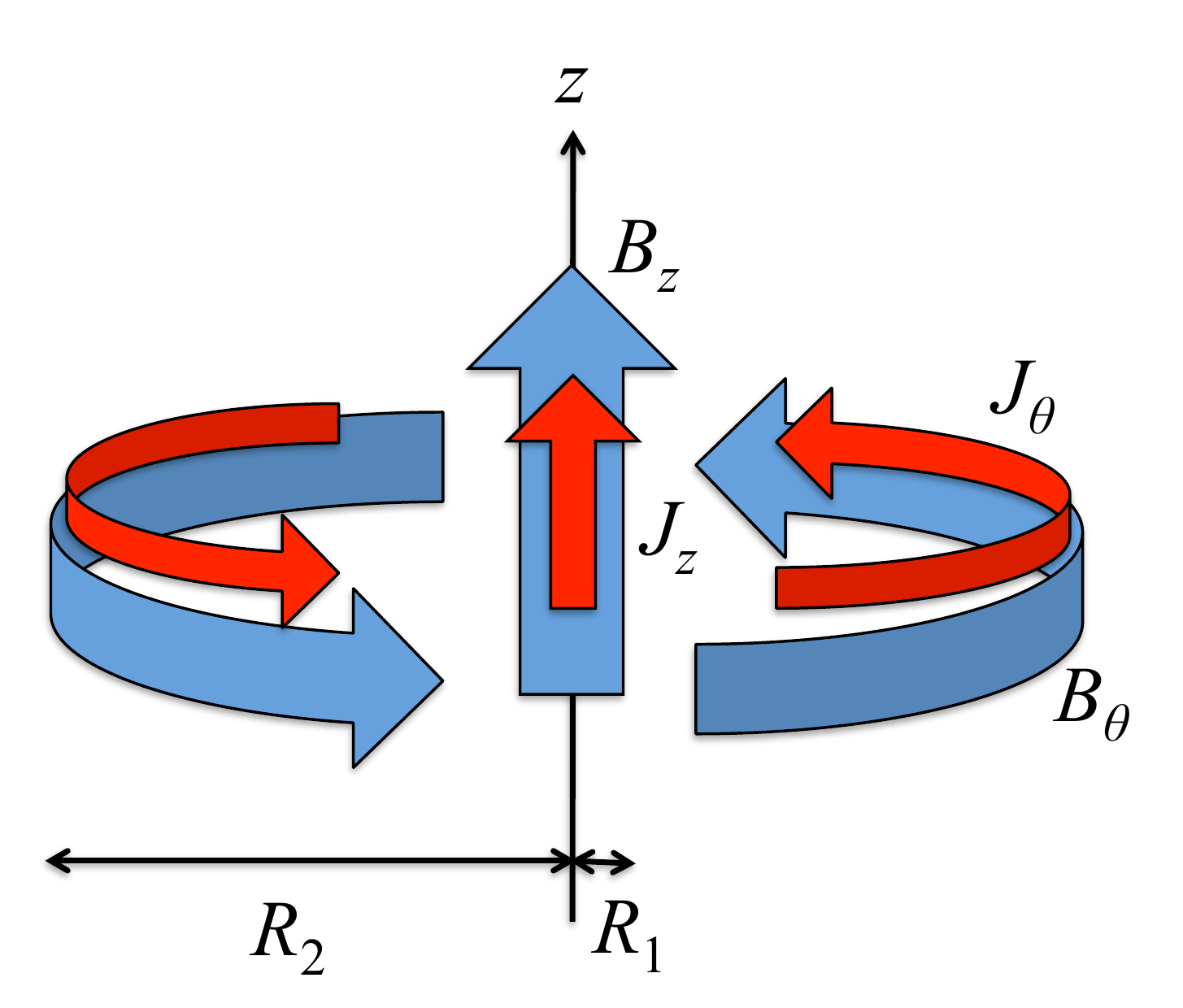}
\end{center}
\vspace{-0.5cm}
\caption{Intuitive picture of the chiral plasma instability.
Blue and red arrows denote magnetic fields and electric currents, respectively.} 
\label{fig:cpi}
\end{figure}

Suppose there is a fluctuation of a magnetic field $\bm B_{\rm in}=B_z\bm e_z$ in 
some finite domain along the $z$ axis with a typical radius $R_1$ (of order the 
wavelength of the magnetic field, $\lambda$), as depicted in Fig.~\ref{fig:cpi}.
Here we use a cylindrical coordinate system characterized by three unit vectors,
$({\bm e_r}, {\bm e_{\theta}}, {\bm e_z})$.
By the chiral magnetic effect, a current density $\bm j_{\rm in} = j_z\bm e_z$ with 
$j_z = \kappa B_z$ is induced. Then, according to Amp\`ere's law, this current density 
leads to a magnetic field around the $z$ axis. At a distance $R$ larger than $R_1$, the magnetic 
field is given by $\bm B(R)=B_{\theta}(R)\bm e_{\theta}$ with $B_{\theta}(R)=\pi R_1^2 j_z/2\pi R$.
Again by the chiral magnetic effect, a current density $\bm j(R) = j_{\theta}(R)\bm e_{\theta}$ 
with $j_{\theta}(R)=\kappa B_{\theta}(R)$ is there.
Amp\`ere's law suggests that the induced current $j_{\theta}(R)$ in the region $R_1<R<R_2$ 
produces a magnetic field along the $z$ axis, $\bm B'_{\rm in} = B'_z \bm e_z$, where
\beq
B'_z=\int_{R_1}^{R_2}j_{\theta}(R) dR = \frac{\kappa^2}{2} R_1^2 B_z \ln \left(\frac{R_2}{R_1} \right).
\eeq
Here the UV and IR cutoffs, $R_1$ and $R_2$, are comparable to $\lambda$, so
the logarithmic factor is just some numerical factor.

Let us first consider the static case and ignore the time dependence of electromagnetic fields.
In this case, the condition $B_z=B'_z$ must be satisfied. This amounts to the condition,
${\bm \nabla} \times {\bm B} = \kappa {\bm B}$ (known as the Beltrami field), 
which is a force-free field because ${\bm j} \times {\bm B}=0$.
In the context of electroweak plasmas, such a static solution was given in 
Refs.~\cite{Rubakov:1985nk, Rubakov:1986am} and is called the Chern-Simons (CS)
wave (see also Refs.~\cite{Khaidukov:2013sja, Manuel:2013zaa} for recent works).
Then one finds that $\lambda_c\sim 1/\kappa$ is the critical value for the static solution. 

If $\lambda > \lambda_c$ $(B'_z>B_z)$, on the other hand, the static situation can no 
longer be sustained and the (electro)magnetic fields grow, and thus, they are unstable.
This is the chiral plasma instability.
In the presence of such an instability, one needs to take into account the time dependence
of electromagnetic fields. A growing magnetic field $B_z(t)$ as a function of time means that, 
according to Faraday's law, an electric field is also induced at a distance $R$: 
$\bm E(R)=E_{\theta}(R)\bm e_{\theta}$ 
with $E_{\theta}(R)<0$. Note that the direction of this electric field $\bm E(R)$ is the \emph{opposite} 
of that of the magnetic field $\bm B(R)$. Now remember the anomaly relation, which connects the 
nonconservation of the chiral charge to the electromagnetic fields: 
$\Delta N_5 = e^2/(2\pi^2) \int d^3{\bm x} \, {\bm E} \cdot {\bm B}$.
That the induced electric and magnetic fields are in the opposite direction means, 
according to the anomaly relation, that the chiral charge $N_5$ (and consequently, $\mu_5$) 
must decrease as a function of time. In this way, the chirality imbalance tends to switch itself 
off, and thereby, the chiral plasma instability is weakened. 

One may repeat a similar argument for non-Abelian plasmas.
In this case, the critical value $1/\lambda_c$ above is the magnetic scale $g^2 T$ for 
$\mu_5 \sim T$, which is much \emph{shorter} than the mean free path, 
$l_{\rm mfp} \sim (g^4 T)^{-1}$, as mentioned above. This is why hydrodynamics cannot
describe the non-Abelian chiral plasma instability. Apparently, such dynamics should be
important for the nonequilibrium evolution of a chiral plasma. For example, one would 
naively expect that the typical time scale of the color fluctuations is modified in the presence
of the chiral plasma instability. In this paper, we describe these systems by a simple effective 
theory at the magnetic scale, and show that the typical time scale of the chiral plasma 
instability is of order $[g^4 T \ln(1/g)]^{-1}$  (which is comparable to the mean free time 
for large angle scattering).

The paper is organized as follows. In Sec.~\ref{sec:bottom-up}, after reviewing the 
conventional Langevin theory for the dynamics of soft gauge fields, we provide a 
physical derivation of the Langevin theory with anomalous effects (the chiral Langevin 
theory). In Sec.~\ref{sec:top-down}, we derive the chiral Langevin theory from the
chiral kinetic theory by integrating out the (semi)hard degrees of freedom. In 
Sec.~\ref{sec:instability}, we apply the chiral Langevin theory and compute the typical 
time scale of the chiral plasma instability. Section~\ref{sec:conclusion} is devoted to 
conclusions.

\section{Physical arguments}
\label{sec:bottom-up}
\subsection{Langevin theory for soft gauge fields without anomalous effects}
We first briefly review the Langevin-type effective theory without anomalous
parity-violating effects that describes the nonperturbative physics at the 
soft scale $g^2 T$. This effective theory was first constructed starting from 
the Boltzmann-Vlasov-Maxwell-type kinetic equation in Ref.~\cite{Bodeker:1998hm}, 
and later a more intuitive derivation was given in Ref.~\cite{Arnold:1998cy}
(see also Ref.~\cite{Litim:1999ns} for another derivation).
We here explain the intuitive argument of Ref.~\cite{Arnold:1998cy}.

We split the gauge fields into those at the soft scale $k \sim g^2 T$ and 
(semi)hard scale $k \gg g^2 T$.
Let us now remember that the dynamics of soft modes is classical.
This is because, for the soft modes $k \sim g^2 T$, the Bose-Einstein factor
\beq
n(k)=\frac{1}{e^{k/T}-1} \simeq \frac{T}{k} \gg 1,
\eeq 
is so large that quantum effects are negligible.
Thus, the soft modes are effectively described by the \emph{classical} 
Yang-Mills equation
\beq
\label{YM_cl}
{\bm D} \times {\bm B} = D_t {\bm E} + {\bm j}_{\rm hard},
\eeq
where ${\bm B} = {\bm D} \times {\bm A}$ and the covariant derivatives $D$
are understood as only involving the soft modes, and ${\bm j}_{\rm hard}$
is the color current consisting of hard modes. Here and below, we suppress 
the color indices of colored quantities, unless there is a possibility of confusion.
In the following, we shall take the $A_0 = 0$ gauge, where the covariant derivative 
$D_t$ reduces to $\d_t$.

In the parity-invariant system (i.e., in the system without a chirality imbalance),
the color current is written in the form of a non-Abelian analogue of the 
Ohmic law
\beq
{\bm j}_{\rm hard} = \sigma_c {\bm E},
\eeq
where $\sigma_c$ is the color conductivity expressed by 
\cite{Arnold:1998cy, Selikhov:1993ns, Heiselberg:1994px}
\beq
\label{color}
\sigma_c = \frac{m_D^2}{3 \gamma} 
\sim \frac{T}{\ln(1/g)},
\eeq
where $m_D$ is the Debye mass and
\beq
\label{gamma}
\gamma = N_c \alpha_g T \ln(1/g)
\eeq
is the damping rate for hard thermal bosons 
with $N_c$ being the number of colors and $\alpha_g = g^2/(4\pi)$.
As will be verified {\it a posteriori}, the typical time scale $\tau$ of Eq.~(\ref{YM_cl}) 
is much larger than $\sigma^{-1}$ [see Eq.~(\ref{tau})], and thus the $D_t {\bm E}$ 
term is negligible compared with the $\sigma_c {\bm E}$ term. 
In the $A_0=0$ gauge, the Yang-Mills equation above reduces to
\beq
\label{Ampere}
\sigma_c \d_t {\bm A} = - {\bm D} \times {\bm B}.
\eeq
However, this equation does not account for the correct 
equilibrium fluctuations of the soft modes. This is because
the soft modes are not only relaxed away from the equilibrium, but also 
excited from thermal noise due to the interactions with hard modes,
the latter of which is missed here. To take into account the latter effect, 
we add the noise term $\zeta$ to have
\beq
\label{YM+noise}
\sigma_c \d_t {\bm A} = - {\bm D} \times {\bm B} + {\bm \zeta}.
\eeq

It is simple to obtain the equation that ${\bm \zeta}$ must satisfy;
according to the fluctuation-dissipation theorem, which connects 
dissipation to thermal noise, ${\bm \zeta}$ must satisfy
\beq
\label{FD}
\langle \zeta^a_i(x) \zeta^b_j(x') \rangle = 2\sigma_c T \delta_{ij}\delta^{ab} 
\delta^{(4)}(x-x'),
\eeq
where $i,j$ and $a,b$ are spatial and color indices, respectively, 
and $\sigma_c$ is the same conductivity as above.

Note that Eq.~(\ref{YM+noise}) may also be written in a more familiar form
of the Langevin-type equation
\beq
\label{Langevin}
\sigma_c \d_t {\bm A} = - {\bm \nabla}_{\bm A} H_{\rm eff}({\bm A}) + {\bm \zeta},
\eeq
where $H_{\rm eff}$ is the three-dimensional effective Hamiltonian given by
\beq
\label{MS}
\qquad H_{\rm eff}({\bm A}) = \frac{1}{2}\int d^3{\bm x}\, {\bm B}^2.
\eeq
It should be remarked that $H_{\rm eff}/T$ is exactly the magnetostatic Yang-Mills
action, which can be obtained by performing the dimensional reduction in the 
$T$ direction and by integrating out the electrostatic field $A_0$ \cite{Braaten:1995jr}.

Once the effective theory (\ref{YM+noise}) is obtained, one can read off the 
typical time scale of the dynamics as
\beq
\label{tau}
\tau \sim \frac{\sigma_c}{k^2} \sim \frac{1}{g^4 T \ln(1/g)},
\eeq
where Eq.~(\ref{color}) is used.
Therefore, $\sigma_c$ in Eq.~(\ref{color}) is much larger than 
$gA \sim g^2 T$ and $\tau^{-1}$, and the approximation to obtain 
Eq.~(\ref{Ampere}) is indeed justified.
This result was first obtained in Ref.~\cite{Bodeker:1998hm}.

\subsection{Intuitive derivation of the chiral Langevin theory}
\label{sec:intuitive}
We now consider the system with a chirality imbalance parametrized 
by the chiral chemical potential $\mu_5 = (\mu_R - \mu_L)/2$.
In this case, the color current consists of not only the usual Ohmic current, 
but also the anomalous current,
\beq
{\bm j}_{\rm hard} = \sigma_c {\bm E} + \sigma_{\rm anom} {\bm B}. 
\eeq
Looking at this relation, one might think that the current proportional to ${\bm B}$ 
must be prohibited due to the parity; while the (non-Abelian) Ohmic law is consistent 
with parity (as ${\bm j}_{\rm hard}$ is parity odd and ${\bm E}$ is parity odd), 
the anomalous current is not (as ${\bm B}$ is parity even). However,
the anomalous current with $\sigma_{\rm anom} \propto \mu_5$ is 
in fact consistent with parity as $\mu_5$ is also parity odd.
As shown below, one finds that $\sigma_{\rm anom}$ is given by
\beq
\sigma_{\rm anom} = \frac{N_f g^2 \mu_5}{4\pi^2},
\eeq
where $N_f$ is the number of flavors.
Note that the usual conductivity $\sigma_c$ is dissipative
(as ${\bm j}_{\rm hard}$ is time-reversal odd and ${\bm E}$ is time-reversal even),
while the anomalous conductivity $\sigma_{\rm anom}$ is dissipationless
(as ${\bm B}$ is time-reversal odd). 
Note also that the prefactor $1/4\pi^2$ (per each flavor) is different
from that of the Abelian chiral magnetic effect \cite{Vilenkin:1980fu,Nielsen:1983rb,Alekseev:1998ds,Son:2004tq,Fukushima:2008xe}, 
$1/2\pi^2$, due to the normalization of the non-Abelian charges, 
$\tr[t^a t^b]=(1/2)\delta^{ab}$.

From Eq.~(\ref{tau}), the magnitude of the electric field $E$ is
\beq
\label{E}
E \sim \tau^{-1} A \sim g^5 T^2 \ln(1/g),
\eeq
and, combined with Eq.~(\ref{color}), we have
\beq
j_{\rm ohm} \sim g^5 T^3.
\eeq
On the other hand, the anomalous current can be evaluated as
\beq
j_{\rm anom} \sim g^5 \mu_5 T^2,
\eeq
where Eq.~(\ref{scale}) is used. 
Therefore, $j_{\rm ohm} \sim j_{\rm anom}$ for $\mu_5 \sim T$.

Let us now turn to the intuitive derivation of the Langevin theory for this system. 
Our argument is based on the special nature of the anomalous current that 
it is \emph{topological} and has nothing to do with collisional effects 
(i.e., dissipations or fluctuations).
This may also be understood from the viewpoint of symmetries: 
$\sigma_{\rm anom}$ is time-reversal invariant, while collisional effects are 
not, and they are not related to each other. 
This property leads us to expect that the following replacement would be 
sufficient to obtain the desired Langevin equation:
\beq
{\bm j}_{\rm ohm} \rightarrow {\bm j}_{\rm ohm} + {\bm j}_{\rm anom},
\eeq
with the rest parts of the Langevin equation concerning 
hard modes, which are \emph{dissipative}, unchanged.
It will turn out in Sec.~\ref{sec:top-down} that this hand-waving argument gives 
the correct answer. The resulting effective theory, which we shall call
the chiral Langevin theory, is given by
\beq
\label{chiral_Langevin}
\sigma_c \d_t {\bm A} = - {\bm D} \times {\bm B} +  
\frac{N_f g^2\mu_5}{4\pi^2} {\bm B} + {\bm \zeta},
\eeq
where ${\bm \zeta}$ again satisfies Eq.~(\ref{FD}).

In terms of the effective Hamiltonian $H_{\rm eff}$ in Eq.~(\ref{Langevin}), 
the above modification corresponds to the following modified Hamiltonian:
\begin{gather}
H_{\rm eff}({\bm A}) = \int d^3{\bm x} \,
\left(\frac{1}{2}{\bm B}^2 + 2 N_f \mu_5 n_{\rm CS} \right), \nonumber \\ 
n_{\rm CS} = \frac{g^2}{32\pi^2} 
\epsilon_{ijk}\left(F^a_{ij}A^a_k - \frac{g}{3}f_{abc}A^a_i A^b_j A^c_k \right).
\label{CS}
\end{gather}
The second term in $H_{\rm eff}$ is the induced Chern-Simons term at finite $\mu_5$, 
which was derived by integrating out fermion degrees of freedom in 
Refs.~\cite{Redlich:1984md, Tsokos:1985ny, Niemi:1985ir} 
and was studied in the context of the electroweak theory in 
Refs.~\cite{Rubakov:1985nk, Rubakov:1986am}.
Note that the Chern-Simons term should not receive any perturbative or nonperturbative
corrections because it has a topological origin; as we will see below, it originates from 
the Berry curvature in the kinetic theory.
In the presence of $\mu_5$, the magnetostatic Yang-Mills action in 
Ref.~\cite{Braaten:1995jr} must be modified to the form of Eq.~(\ref{CS}).
Note also that the modified Yang-Mills action $H_{\rm eff}/T$ with Eq.~(\ref{CS})
has only one scale $g^2 T$ for $\mu_5 \sim T$ (for which $j_{\rm ohm} \sim j_{\rm anom}$ 
as mentioned above).

In the electroweak theories considered in Refs.~\cite{Redlich:1984md, Tsokos:1985ny, Niemi:1985ir,
Rubakov:1985nk, Rubakov:1986am}, 
dissipative effects are not included. We stress that our Langevin-type equation in 
Eq.~(\ref{Langevin}) together with Eq.~(\ref{CS}) is a more (and the most) complete framework 
taking into account the effects of the dissipation and thermal fluctuations systematically.
It is thus appropriate to be applied to study the dynamical evolution of non-Abelian chiral plasmas.
Converting this chiral Langevin equation with Gaussian white noise into a Fokker-Planck equation, 
one can check that it reproduces the correct ``equilibrium" distribution $e^{-H_{\rm eff}/T}$ 
(see Appendix~\ref{sec:FP}).%
\footnote{
One might suspect that $e^{-H_{\rm eff}/T}$ is not the equilibrium 
distribution in the usual sense, because, as we argued, the system with $\mu_5$ 
has a chiral plasma instability which tends to reduce $\mu_5$ dynamically. 
However, it will be justified {\it a posteriori} that $\mu_5$ changes very slowly compared 
with the typical time scale of the system for $g \ll 1$ (see Sec.~\ref{sec:instability}), so
$e^{-H_{\rm eff}/T}$ can be approximately regarded as the ``equilibrium" distribution.
}

In Sec.~\ref{sec:top-down}, we provide a more detailed derivation of the chiral 
Langevin theory, starting from the chiral kinetic theory and integrating out (semi)hard 
degrees of freedom. As found in Refs.~\cite{Son:2012wh, Stephanov:2012ki, Son:2012zy, Chen:2012ca}, 
anomalous parity-violating effects are taken into account by introducing Berry 
curvature corrections \cite{Berry}, a notion widely applied in condensed matter 
physics \cite{Xiao:2010, Volovik}. Because the kinetic theory with Berry curvature 
corrections can be derived from the underlying quantum field theories \cite{Son:2012zy, Chen:2012ca}, 
this will complete the derivation of the chiral Langevin theory from quantum field theories.

\subsection{Evolution equation for the chiral charge}
\label{sec:mu5}
Here we comment on the chiral chemical potential $\mu_5$.
One might suspect that $\mu_5$ cannot be introduced as the usual chemical 
potential associated with a conserved charge, because the chiral charge 
is not a conserved quantity due to the axial anomaly:
\beq
\label{anomaly}
\d_{\mu} j^{\mu 5} = \frac{N_f g^2}{4\pi^2} {\bm E} \cdot {\bm B}.
\eeq
Here $j^{\mu 5} = (n_5, {\bm j}^5)$ is the (color- and flavor-singlet) axial current and 
$N_f$ is the number of flavors.
Nonetheless, one may introduce $\mu_5$ as an external parameter if the time scale of 
its variation is slow enough compared with the typical time scale of the system.%
\footnote{
Remember that the baryon chemical potential is also introduced although 
baryon charge is \emph{not} strictly a conserved charge due to quantum anomalies 
in the Standard Model. The reason why it makes sense to do so is because the time 
scale of baryon-number-changing processes, such as the proton lifetime $\tau_p$, 
is too large (much larger than the age of the Universe), and during the time scale 
much less than $\tau_p$, the baryon charge is regarded as conserved.
}
This will be actually justified {\it a posteriori} for $g \ll 1$ in Sec.~\ref{sec:instability}.
In general, one needs to treat $\mu_5$ as a \emph{dynamical} variable which also 
evolves as a function of time.

Let us derive the evolution equation for $\mu_5(t, {\bm x})$. As diffusion of axial charge 
occurs at the length scale much larger than $(g^2 T)^{-1}$ \cite{Arnold:2000dr}, we can 
simply set ${\bm j}^5 = 0$ at the soft scale (in the absence of electromagnetic fields). 
Then we get
\beq
\label{eq:n_5}
\d_t n_5 = \frac{N_f g^2}{4\pi^2} {\bm E} \cdot {\bm B}.
\eeq
In equilibrium, the axial number density $n_5$ is expressed by the 
chiral chemical potential $\mu_5$ via
\beq
\label{n_5}
n_5 \equiv n_R - n_L 
= N_c N_f \left[\frac{1}{3 \pi^2} (\mu_5^3 + 3 \mu_5 \mu^2) + \frac{1}{3}T^2 \mu_5\right],
\eeq
which we assume to remain valid, as the system is close to equilibrium.

In summary, the dynamical variables in our chiral Langevin theory
are ${\bm A}(t, {\bm x})$ and $\mu_5(t,{\bm x})$, whose evolutions are 
described by Eqs.~(\ref{chiral_Langevin}) and (\ref{eq:n_5}).

\section{From chiral kinetic theory to the chiral Langevin theory}
\label{sec:top-down}
In this section, we derive the chiral Langevin theory from the chiral kinetic theory 
for a non-Abelian plasma. 

\subsection{Chiral kinetic theory}
Let us first recall the (collisionless) Boltzmann equation for an Abelian plasma
(for which we use the coupling constant $e$ instead of $g$).
It is formulated in terms of the single-particle phase-space 
distribution function of hard particles, $n(t, {\bm x}, {\bm p})$. 
According to Liouville's theorem, it follows that
\beq
\frac{dn}{dt} = 0.
\eeq
Noting that $n$ is a function of $t$, ${\bm x}$, and ${\bm p}$,
the left-hand side of the above equation can be expanded as
\beq
\frac{dn}{dt} = \left(\d_t + \dot {\bm x} \cdot  {\bm \nabla}_{\! \! x} + 
\dot {\bm p} \cdot  {\bm \nabla}_{\! \! p} \right)n.
\eeq
Usually, the classical equations of motion for hard particles are given by
\begin{subequations}
\begin{align}
\label{EOM_conv_x}
\dot {\bm x} & = {\bm v}, \\
\label{EOM_conv_p}
\dot{\bm p} & = e({\bm E} + \dot {\bm x} \times {\bm B}),
\end{align}
\end{subequations}
where ${\bm v} = d \epsilon_{\bm p}/d {\bm p} = {\bm p}/p \equiv \hat {\bm p}$ 
is the group velocity of the plasma particles. (We here defined  $p \equiv |{\bm p}|$.)
Then one obtains
\beq
\label{Boltzmann}
\d_t n_{\bm p} + {\bm v} \cdot {\bm \nabla}_{\! \! x} n_{\bm p} 
+ e({\bm E} + \dot {\bm x} \times {\bm B}) \cdot {\bm \nabla}_{\! \! p}n_{\bm p} = 0,
\eeq
which is the conventional Boltzmann equation.
The electric current is given by
\beq
{\bm j} \equiv \int \frac{d^3{\bm p}}{(2\pi)^3}\, \dot {\bm x} n_{\bm p} = 
\int \frac{d^3{\bm p}}{(2\pi)^3}\, {\bm v} n_{\bm p}.
\eeq

For chiral fermions, the equations of motion are modified 
by the effects of a Berry curvature:%
\footnote{
This modification may be understood as follows.
First, remember the Hamiltonian for a spin in a magnetic field,
$H_{\rm spin} = {\bm \sigma} \cdot {\bm B}$,
which was considered in the original paper by Berry \cite{Berry}. 
In this case, the Berry curvature in the ${\bm B}$ space is found to be 
${\bm \Omega}_{\bm B} = \pm \frac{\bm B}{2|{\bm B}|^3}$
in the magnetic-field space $(B_x, B_y, B_z)$,
where the signs $\pm$ correspond to the spin polarizations
along the direction of the magnetic field \cite{Berry}. In our case, the Hamiltonian 
for a chiral fermion is $H_{\rm chiral} = {\bm \sigma} \cdot {\bm p}$,
which is equivalent to the Hamiltonian above if we replace
${\bm B}$ by ${\bm p}$; the corresponding Berry curvature in the
momentum space is given by 
${\bm \Omega}_{\bm p} = \pm \frac{\bm p}{2|{\bm p}|^3}$,
where the signs $\pm$ correspond to the chiralities of fermions.
}
the correct equations of motion are given by \cite{SundaramNiu}
\begin{subequations}
\begin{align}
\label{EOM_chiral_x}
\dot {\bm x} & =\frac{\d \epsilon_{\bm p}}{\d {\bm p}} + 
\dot {\bm p} \times {\bm \Omega}_{\bm p}, \\
\label{EOM_chiral_p}
\dot{\bm p} & = e({\bm E} + \dot {\bm x} \times {\bm B}) - 
\frac{\d \epsilon_{\bm p}}{\d {\bm x}}.
\end{align}
\end{subequations}
Here ${\bm \Omega}_{\bm p} = \pm {\bm p}/(2p^3)$
is the Berry curvature and the signs $\pm$ correspond to the right- and left-handed 
fermions \cite{Volovik, Son:2012wh, Stephanov:2012ki, Son:2012zy};
$\epsilon_{\bm p}$ is the energy of the fermion as a function of ${\bm p}$ and
its form is determined from the Lorentz covariance as \cite{Son:2012zy}
\beq
\label{energy}
\epsilon_{\bm p} = p(1 - e {\bm B} \cdot {\bm \Omega}_{\bm p}).
\eeq
Physically, the second term in Eq.~(\ref{energy}) stands for the magnetic 
moment of chiral fermions in a magnetic field (the Zeeman effect). 
Note that Eqs.~(\ref{EOM_chiral_x}) and (\ref{EOM_chiral_p}) reduce to 
Eqs.~(\ref{EOM_conv_x}) and (\ref{EOM_conv_p}) in the absence of 
Berry curvature corrections as they should.

By solving Eqs.~(\ref{EOM_chiral_x}) and (\ref{EOM_chiral_p}) for 
$\dot {\bm x}$ and $\dot {\bm p}$, one finds
\begin{align}
\label{xdot2}
(1 + e {\bm B} \cdot {\bm \Omega}_{\bm p})
\dot {\bm x} & = \tilde {\bm v} + e \tilde {\bm E} \times {\bm \Omega}_{\bm p}
+ (\tilde {\bm v} \cdot {\bm \Omega}_{\bm p}) e {\bm B}
\\
\label{pdot2}
(1 + e {\bm B} \cdot {\bm \Omega}_{\bm p})
\dot {\bm p} & = e \tilde {\bm E} + \tilde {\bm v} \times e {\bm B} 
+ e^2 (\tilde {\bm E} \cdot {\bm B}) {\bm \Omega}_{\bm p},
\end{align}
where $\tilde {\bm v} \equiv \d \epsilon_{\bm p}/\d {\bm p}$ and 
$\tilde {\bm E} \equiv {\bm E} - {{\bm \nabla} (p{\bm B} \cdot {\bm \Omega}_{\bm p})}$.
Then the modified Boltzmann equation with Berry curvature corrections is 
\cite{Son:2012zy} (see also Refs.~\cite{Duval:2005, Son:2012bg, Stephanov:2012ki})
\begin{align}
\label{Boltzmann_Berry}
(1 + e {\bm B} \cdot {\bm \Omega}_{\bm p})\d_t n_{\bm p}
+ \left(\tilde {\bm v} + e \tilde {\bm E} \times {\bm \Omega}_{\bm p}
+ (\tilde {\bm v} \cdot {\bm \Omega}_{\bm p}) e {\bm B} \right) \cdot {\bm \nabla}_{\! \! x} n_{\bm p}
\nonumber \\
+ \left(e \tilde {\bm E} + \tilde {\bm v} \times e {\bm B} + e^2 (\tilde {\bm E} \cdot {\bm B}) 
{\bm \Omega}_{\bm p}\right) \cdot {\bm \nabla}_{\! \! p} n_{\bm p} = 0.
\end{align}
The modification of the equations of motion means that the phase-space measure is 
also modified from $d\Gamma = d{\bm x} d{\bm p}$ to \cite{Xiao:2005}
\beq
\label{phase-space}
d\Gamma = \sqrt{\omega} d{\bm x} d{\bm p} \equiv
(1 + e {\bm B} \cdot {\bm \Omega}_{\bm p}) d{\bm x} d{\bm p}.
\eeq
As a result, the expression for the electric current is 
\cite{Son:2012wh, Stephanov:2012ki, Son:2012zy}
\begin{align}
\label{j}
{\bm j} 
= -e\int \frac{d^3{\bm p}}{(2\pi)^3}\, \left[\epsilon_{\bm p} {\bm \nabla}_{\! \! p}n_{\bm p}
+ e\left({\bm \Omega}_{\bm p} \cdot {\bm \nabla}_{\! \! p}n_{\bm p} \right) \epsilon_{\bm p} {\bm B}
+ \epsilon_{\bm p} {\bm \Omega}_{\bm p} \times {\bm \nabla}_{\! \! x} n_{\bm p} \right] 
+ {e^2\bm E} \times {\bm \sigma}, 
\end{align}
where 
\beq
\label{sigma}
{\bm \sigma} = \int \frac{d^3{\bm p}}{(2\pi)^3}\, {\bm \Omega}_{\bm p} n_{\bm p}.
\eeq

\subsection{Linearized Boltzmann equation}
Let us now linearize the Boltzmann equation above. 
We define the ``scalar potential" $W=W(x,{\bm v})$ such that 
$n(x, {\bm p}) = n^{\rm eq}(\epsilon_{\bm p} - e W)$. Then the distribution 
function $n(x, {\bm p})$ can be expanded up to linear terms in $W$.
Using Eq.~(\ref{energy}), it follows that
\beq
\label{n_linear}
n(x, {\bm p}) 
\simeq n^{\rm eq}_{\bm p} - \frac{d n^{\rm eq}_{\bm p}}{d p} 
e(W + p{\bm B} \cdot {\bm \Omega}_{\bm p}),
\eeq 
where $n^{\rm eq}_{\bm p}$ is the equilibrium distribution function, 
$n^{\rm eq}_{\bm p} = (e^{(p-\mu)/T}+1)^{-1}$. 
In terms of the deviation $\delta n$ from equilibrium 
(including the Zeeman effect in a magnetic field), 
it is also written as
\beq
\delta n \equiv n(x,{\bm p}) - n^{\rm eq}(\epsilon_{\bm p})
\simeq -e W \frac{d n^{\rm eq}_{\bm p}}{d p}.
\eeq

The linearized Boltzmann equation takes the following form:
\begin{align}
\label{linearized_Boltzmann1}
v \cdot \d_x W_{\rm R, L}(x, {\bm v}) = {\bm v} \cdot {\bm E}(x) \mp \frac{{\bm v}}{2p} \cdot \d_t {\bm B}(x),
\end{align}
for right- and left-handed fermions, 
where $v \cdot \d_x \equiv \d_t + {\bm v} \cdot {\bm \nabla}_{\! \! x}$.
Similarly, the linearized Boltzmann equation for antiparticles is
\begin{align}
\label{linearized_Boltzmann2}
v \cdot \d_x \overline W_{\rm R,L}(x, {\bm v}) 
= - {\bm v} \cdot {\bm E}(x) \mp \frac{{\bm v}}{2p} \cdot \d_t {\bm B}(x).
\end{align}
Note that antiparticles of right-(left-)handed fermions are left-(right-)handed.
It is clear from Eqs.~\eqref{linearized_Boltzmann1} and \eqref{linearized_Boltzmann2} that 
the electric field $\bm E$ couples to a scalar potential carrying an electric charge, 
$W_+\equiv (W_{\rm R}+W_{\rm L}-\overline W_{\rm R}-\overline W_{\rm L})/4$, 
while the magnetic field $\bm B$ couples to that carrying a magnetic moment, 
$W_-\equiv (p/2)(W_{\rm R}+\overline W_{\rm L} -W_{\rm L}-\overline W_{\rm R})$.
The linearized Boltzmann equations for the scalar potentials $W_+$ \cite{Blaizot:2001nr}
and $W_-$ \cite{Son:2012zy, Manuel:2013zaa} are obtained as 
\begin{align}
\label{linearized_Boltzmann}
v \cdot \d_x W_+(x, {\bm v}) &= {\bm v} \cdot {\bm E}(x), \\
v \cdot \d_x W_-(x, {\bm v}) &= - {\bm v} \cdot \d_t {\bm B}(x).
\end{align}

Considering the charge conjugation symmetry (for the energy shift of 
fermions in the electromagnetic fields), it is natural to take 
$W_{\rm R}=-\overline W_{\rm R}$ and $W_{\rm L}=-\overline W_{\rm L}$.
Summing over $W_{\rm R,L}$ and $\overline W_{\rm R,L}$ in Eq.~(\ref{j}) and 
performing the integral over $p \equiv |{\bm p}|$, 
we have \cite{Manuel:2013zaa}
\beq
{\bm j}=m_D^2\int_{\bm v} {\bm v} W_+ + 
\frac{\mu_5e^2}{2\pi^2}\int_{\bm v} 
\left({\bm v} W_- + {\bm B} - {\bm v} \times {\bm \nabla} W_+\right),
\eeq
where $\int_{\bm v} \equiv \int {d \Omega}/(4\pi)$ is the angular integral and
\beq
m_D^2 \equiv -e^2 \int \frac{d^3 {\bm p}}{(2\pi)^3}\, \frac{d n^{\rm eq}_{\bm p}}{d p}
= e^2 \left(\frac{T^2}{3} + \frac{\mu^2 + \mu_5^2}{\pi^2}\right).
\eeq

\subsection{Linearized non-Abelian Boltzmann-Vlasov equation}
We shall generalize this set of linearized Boltzmann equation to the case of a  non-Abelian 
plasma with $N_c$ colors and $N_f$ flavors. 
Here we do not attempt to derive it from a first-principles Wigner function approach 
or the Kadanoff-Baym formalism (see, e.g., Ref.~\cite{Blaizot:2001nr} for a review), 
but rather we try to derive it on physical grounds. Such a microscopic derivation should 
be doable by generalizing the argument of Ref.~\cite{Son:2012zy} to non-Abelian gauge fields.

First, notice that the color current is carried by both fermions and bosons
in non-Abelian plasmas (e.g., quarks and gluons in QCD) while the electric current is 
carried only by fermions in Abelian plasmas  (e.g., electrons in QED). So one needs to introduce 
the single particle phase-space distribution functions both for fermions and bosons, which
will be denoted as $n^a$ and $N^a$. 
Similarly to the Abelian plasma, we parametrize
\begin{align}
C_F \delta n^a(x, {\bm p}) = -g W_F^a \frac{d n^{{\rm eq}}_F}{d p}, \qquad
C_B \delta N^a(x, {\bm p}) = -g W_B^a \frac{d n^{{\rm eq}}_B}{d p}.
\end{align}
Here $n^{\rm eq}_{F,B}$ is the equilibrium Fermi/Bose distribution
function (which is colorless), and $W_B = W_B^a t^a$ and 
$W_F = W_F^a t^a$, where $t^a$ ($a=1,2, \cdots, {N_c}^2 -1$) are
the generators of fundamental representations of $\SU(N_c)$ with
the normalization such that $\tr(t^a t^b) = (1/2) \delta^{ab}$.
The factors $C_{F,B}$ reflect the fundamental and adjoint representations 
of the $\SU(N_c)$ group for fermions and bosons, and are given by 
$C_F=1/2$ and $C_B=N_c$, respectively.

Second, note that the motion of chiral fermions is affected by the Berry curvature, 
while that of gauge bosons is not. For gauge bosons, $W_B$ thus consists only of 
the parity-even part $W_{B+}$ and the linearized Boltzmann equation is known as 
\cite{Blaizot:2001nr}
\beq
[v \cdot D, W_{B+}(x, {\bm v})] = {\bm v} \cdot {\bm E}(x), 
\eeq
where $v \cdot D = v^{\mu} D_{\mu}$ and $D_{\mu} = \d_{\mu} - ig A_{\mu}$ is the 
covariant derivative. Again, the color indices are suppressed for simplicity.
The color current is
\beq
{j}^{\mu a}_B = m_{D}^2 \int_{\bm v} {v}^{\mu} W^a_{B+}(x, {\bm v}).
\eeq

For chiral fermions, $W_F$ consists of parity-even and parity-odd parts,
$W_{F+}$ and $W_F-$, similarly to the Abelian case above, and the 
linearized Boltzmann equation is given by a non-Abelian generalization
of Eq.~(\ref{linearized_Boltzmann}):
\begin{align}
[v \cdot D, W_{F+} (x, {\bm v})] &= {\bm v} \cdot {\bm E}(x), \\
[v \cdot D, W_{F-} (x, {\bm v})] &= - {\bm v} \cdot D_t {\bm B}(x).
\end{align}
The color current reads
\begin{align}
n^a_F &= m_D^2\int_{\bm v} W^a_{F+}, \\
{\bm j}^a_F &= m_D^2\int_{\bm v} {\bm v} W^a_{F+}  + 
\frac{N_f g^2 \mu_5}{4\pi^2}\int_{\bm v} 
\left[{\bm v} W^a_{F-} + {\bm B}^a - {\bm v} \times ({\bm D} W_{F+})^a \right].
\label{j_W}
\end{align}
They must be solved together with the Yang-Mills equation
\beq
\label{YM}
[D_{\nu}, F^{\nu \mu}] = j^{\mu}, \qquad  j^{\mu a} = j^{\mu a}_B + j^{\mu a}_F,
\eeq
which describes the evolution of gauge fields.

These equations can be further simplified by considering the counting in $g$.
Observe in Eq.~(\ref{j_W}) that, for the semihard and soft modes, 
the final term $\sim g^2 \mu_5 k W_{F+} \lesssim g^3 T^2 W_{F+}$ is much smaller 
than the first term $\sim m_D^2 W_{F+} \sim g^2 T^2 W_{F+}$ and is negligible. 
Then, the parity-even terms for fermions and bosons are combined into 
$W_+ \equiv W_{B+} + W_{F+}$, so that 
(hereafter $W_{F-}$ will be denoted as $W_{-}$, suppressing the subscript  ``$F$")
\begin{align}
\label{linearized_Boltzmann_even}
[v \cdot D, W_{+} (x, {\bm v})] &= {\bm v} \cdot {\bm E}(x), \\
\label{linearized_Boltzmann_odd}
[v \cdot D, W_{-} (x, {\bm v})] &= - {\bm v} \cdot D_t {\bm B}(x), \\
[D_{\nu}, F^{\nu i}] &= m_D^2\int_{\bm v} {\bm v} W_{+}  + 
\frac{N_f g^2 \mu_5}{4\pi^2}\int_{\bm v} {\bm v} W_{-} + \frac{N_f g^2 \mu_5}{4\pi^2}{\bm B}
\end{align}
with
\beq
m_D^2 = (N_f + 2N_c)\frac{g^2 T^2}{6} + N_f \frac{g^2 (\mu^2 + \mu_5^2)}{2\pi^2}.
\eeq
This set of equations constitutes the effective Boltzmann-Vlasov equation with 
anomalous parity-violating effects.

\subsection{Chiral Langevin theory}
We now integrate out semihard degrees of freedom with $k \gg g^2T$
in the Boltzmann-Vlasov equation to derive the chiral Langevin equation. 
The case without the anomalous parity-violating effects was previously
worked out in Ref.~\cite{Bodeker:1998hm} 
(see also Ref.~\cite{Arnold:1998cy}). We shall  follow their procedure 
and extend it to the case with parity-violating effects below.

Let us first summarize the procedure:
\begin{enumerate}
\item{We decompose the fields into semihard and soft modes, and obtain
their equations of motion.}
\item{We explicitly solve the equation of motion for the semihard modes in terms of 
the fields involving soft modes (defined as $h_{\pm}$ below).}
\item{We substitute the above solutions into the expressions for the fluctuations
(defined as $\xi_{\pm}$ below) and take the statistical average thereof.}
\item{By integrating out $W_{\pm}$, we express the color current in terms of the 
gauge fields (${\bm E}$ and $\bm B$) and the fluctuation ${\bm \zeta}$ alone.}
\item{Combined with the Yang-Mills equation, we obtain the Langevin-type equation
for soft fields (\ref{chiral_Langevin}). The fluctuation-dissipation theorem (\ref{FD}) 
also follows from the explicit expression for ${\bm \zeta}$.}
\end{enumerate}

From now on, we shall proceed to take the above steps one by one. 
We work in the $A_0 = 0$ gauge.
\\ 

[{\it Step 1}]: We decompose the fields ${\bm A}$, ${\bm E}$, ${\bm B}$, and $W_{\pm}$ into
\begin{align}
{\bm A} \rightarrow \tilde {\bm A} + {\bm a}, \qquad
{\bm E} \rightarrow \tilde {\bm E} + {\bm e}, \qquad
{\bm B} \rightarrow \tilde {\bm B} + {\bm b}, \qquad
W_{\pm} \rightarrow \tilde W_{\pm} + w_{\pm},
\end{align}
where $\tilde {\bm A}$, $\tilde {\bm E}$, $\tilde {\bm B}$, and $\tilde W_{\pm}$ 
are soft components (components with momenta $k<\mu$, where $g^2 T \ll \mu \ll g T$) 
and ${\bm a}$, ${\bm e}$, ${\bm b}$, and $w$ are semihard components 
(components with $k>\mu$). For simplicity, we shall rename $\tilde {\bm A}$, 
$\tilde {\bm E}$, $\tilde {\bm B}$, and $\tilde W_{\pm}$ as ${\bm A}$, ${\bm E}$, 
${\bm B}$, and $W_{\pm}$ in the following.
We will ignore the interactions between semihard and soft modes which correspond to 
nonhard thermal loop (or bare) vertices, because they are subleading in $g$ 
\cite{Bodeker:1998hm}.
Substituting these expressions into Eqs.~(\ref{linearized_Boltzmann_even}) 
and (\ref{linearized_Boltzmann_odd}), the equations for the soft fields read 
\begin{align}
\label{W+}
[v \cdot D, W_{+} (x, {\bm v})] &= {\bm v} \cdot {\bm E}(x) + \xi_+(x,{\bm v}), \\
\label{W-}
[v \cdot D, W_{-} (x, {\bm v})] &= - {\bm v} \cdot \d_t {\bm B}(x) + \xi_-(x,{\bm v}), \\
\label{EOM_A}
[D_{\nu}, F^{\nu i}] &= m_D^2\int_{\bm v} {\bm v} W_{+}  + 
\frac{N_f g^2 \mu_5}{4\pi^2}\int_{\bm v} {\bm v} W_{-} + \frac{N_f g^2 \mu_5}{4\pi^2}{\bm B},
\end{align}
where
\begin{subequations}
\label{xi}
\begin{eqnarray}
\xi_{+}(x,{\bm v}) &\equiv& -ig[{\bm v} \cdot {\bm a}(x), w_{+}(x,{\bm v})]_{\rm soft},  \\
\xi_{-}(x,{\bm v}) &\equiv& -ig[{\bm v} \cdot {\bm a}(x), w_{-}(x,{\bm v})]_{\rm soft}
-\frac{ig}{2}\epsilon_{ijk}v^i\d_t [a^j,a^k]_{\rm soft}.
\end{eqnarray}
\end{subequations}
The subscript ``soft" stands for the soft components with $k<\mu$.

We then write down equations of motion for the semihard modes.
For semihard modes with momentum $k \sim gT$ 
(and with amplitude $a \sim g^{1/2}T$), the linear term, $ka$, dominates over 
the nonlinear term, $ga^2$, so we can ignore the latter. 
Then the equations of motion read 
\begin{align}
\label{BV_w+}
v \cdot \d w_+ &= {\bm v} \cdot {\bm e} + h_+, \\
\label{BV_w-}
v \cdot \d w_- &= -{\bm v} \cdot \d_t{\bm b} + h_-, \\
\label{EOM_a}
\d^2 {\bm a} + {\bm \nabla}({\bm \nabla} \cdot {\bm a}) &= m_D^2\int_{\bm v} {\bm v} w_{+}  
+ \frac{N_f g^2 \mu_5}{4\pi^2}\int_{\bm v} {\bm v} w_{-} + \frac{N_f g^2 \mu_5}{4\pi^2}{\bm b}.
\end{align}
where $h_{\pm}$ are the interaction terms involving the soft fields,
\begin{align}
\label{h}
h_{\pm} = -ig\left([{\bm v} \cdot {\bm a}, W_{\pm}] + [{\bm v} \cdot {\bm A}, w_{\pm}] \right).
\end{align}
Note that, for the semihard modes with $k \sim gT$, the final term $\propto g^2 \mu_5 {\bm b}$ 
on the right-hand side of Eq.~(\ref{EOM_a}) is suppressed compared with the kinetic term on 
the left-hand side by a factor of $g$, and we will ignore it in what follows.
\\

[{\it Step 2}]: Below we will concentrate on the transverse part of the gauge field 
defined by $a_T^i(t, {\bm k}) = P^{ij}_T a^j(t, {\bm k})$, which we shall rename ${\bm a}(t, {\bm k})$ 
for simplicity. (Here $P_T^{ij} = \delta^{ij} - \hat k^i \hat k^j$ is the transverse projector 
with the unit vector $\hat k^i = k^i/|{\bm k}|$.)
We here ignore the longitudinal part of the semihard mode, because it receives Debye 
screening and does not contribute to the dynamics of the soft modes
\cite{Bodeker:1998hm, Arnold:1998cy}.
In the $(t, {\bm k})$ space, the equations of motion for semihard modes above 
can be written as
\begin{align}
\label{EOM_w+}
\d_t w_+ (t, {\bm k}, {\bm v})+ i {\bm v} \cdot {\bm k} w_+  (t, {\bm k}, {\bm v})
&= -{\bm v} \cdot \d_t {\bm a}(t, {\bm k}) + h_+(t, {\bm k}, {\bm v}), \\
\label{EOM_w-}
\d_t w_- (t, {\bm k}, {\bm v})+ i {\bm v} \cdot {\bm k} w_-  (t, {\bm k}, {\bm v})
&= - i{\bm v} \cdot ({\bm k} \times \d_t{\bm a}(t, {\bm k})) + h_-(t, {\bm k}, {\bm v}), \\
\label{EOM_aT}
\d_t^2 {\bm a}(t, {\bm k}) + |{\bm k}|^2 {\bm a}(t, {\bm k})
&= m_D^2\int_{\bm v} {\bm v}_T w_{+}(t, {\bm k}, {\bm v}) + 
\frac{N_f g^2 \mu_5}{4\pi^2}\int_{\bm v} {\bm v}_T w_{-}(t, {\bm k}, {\bm v}),
\end{align}
where $v_T^i = P^{ij}_T v^j$. These equations can be solved with the help
of the one-sided Fourier transform. The solutions are given by
(see Appendix \ref{sec:solution})
\begin{subequations}
\label{solutions}
\begin{align}
a^i(K) &= a_{(0)}^i(K) + \int_{\bm v}\Delta_{12}^i(K,{\bm v})
\left[h_+ (K,{\bm v}) + \frac{N_f g^2 \mu_5}{4\pi^2 m_D^2}h_- (K,{\bm v}) \right], \\
w_+(K,{\bm v}) &= w_+^{(0)}(K,{\bm v}) 
+ \int_{\bm v'} \left[ \Delta_{22}(K,{\bm v},{\bm v}') h_+ (K,{\bm v'})
+ \frac{N_f g^2 \mu_5}{4\pi^2 m_D^2} \Delta_{23}(K,{\bm v},{\bm v}') h_-(K,{\bm v'}) \right], \\
w_-(K,{\bm v}) &= w_-^{(0)}(K,{\bm v}) 
+ \int_{\bm v'}\Delta_{32}(K,{\bm v},{\bm v}') h_+ (K,{\bm v'})
+\Delta_{33}(K,\bm v)h_-(K,{\bm v}),
\end{align}
\end{subequations}
where $K^{\mu}=(k_0,{\bm k})$, and $a_{(0)}^i$ and $w_{\pm}^{(0)}$ are the solutions to 
the equations of motion for $h_{\pm}=0$, and they can be expressed by the initial values 
at $t=0$, $a_{\rm in}$ and $w_{\pm}^{\rm in}$, alone.
Here we also defined the propagators
\begin{subequations}
\label{propagators}
\begin{align}
\Delta^{i}_{12}(K,{\bm v}) & = \frac{i m_D^2}{v \cdot K}\Delta_T(K) v_T^i, \\
\Delta_{22}(K,{\bm v},{\bm v}') &= \frac{i}{v\cdot K}\delta^{(S^2)}({\bm v}-{\bm v}') 
- \frac{i m_D^2 k_0{\bm v}_T \cdot {\bm v}_T'}{(v \cdot K)(v' \cdot K)} \Delta_T(K), \\
\Delta_{23}(K,{\bm v},{\bm v}') &= 
- \frac{i m_D^2 k_0{\bm v}_T \cdot {\bm v}_T'}{(v \cdot K)(v' \cdot K)} \Delta_T(K), \\
\Delta_{32}(K,{\bm v},{\bm v}') &= 
\frac{m_D^2 k_0{\bm v}_T \cdot ({\bm k}\times {\bm v}_T')}{(v \cdot K)(v' \cdot K)}\Delta_T(K), \\
\Delta_{33}(K,\bm v) &= \frac{i}{v\cdot K},
\end{align}
\end{subequations}
where $\Delta_T(K)$ is the hard thermal loop resummed propagator,
\begin{align}
\Delta_T(K)=\frac{1}{-K^2 + \Pi_T(K)}, \qquad 
\Pi_T(K) = \frac{1}{2}m_D^2 k_0 \int_{\bm v}\frac{{\bm v}_T^2}{v \cdot K},
\end{align}
and $\delta^{(S^2)}$ is the delta function on the two-dimensional sphere $S^2$ such that
\begin{align}
\int_{{\bm v}'} f({\bm v}') \delta^{(S^2)}({\bm v}-{\bm v}')  = f({\bm v}).
\end{align}

Now remember that $h_{\pm}$ themselves are expressed by ${\bm a}$ and $w_{\pm}$ 
[see Eq.~(\ref{h})]. This means one can solve these equations iteratively;
more specifically, inserting ${\bm a}^{(0)}$ and $w_{\pm}^{(0)}$ into Eq.~(\ref{h}) yields 
$h^{(1)}_{\pm}$, and inserting this $h^{(1)}_{\pm}$ into the above equations gives
${\bm a}^{(1)}$ and $w_{\pm}^{(1)}$. Repeating this procedure, one has the 
expansions,
\begin{align}
{\bm a} &= {\bm a}^{(0)} + {\bm a}^{(1)} + {\bm a}^{(2)} + \cdots, \\
w_{\pm} &= w_{\pm}^{(0)} + w_{\pm}^{(1)} + w_{\pm}^{(2)} + \cdots, \\
h_{\pm} &= h_{\pm}^{(1)} + h_{\pm}^{(2)} + h_{\pm}^{(3)} + \cdots,
\end{align}
where the expansion parameter in these expansions is the amplitude of
the soft field $W$ and/or ${\bm A}$, and the indices ``($n$)" 
($n=0,1,2,\cdots$) represent the $n$th order in soft fields.
Note that the expansion parameter is \emph{not} the coupling constant $g$ itself.
Note also that the expansion of $h_{\pm}$ starts from the first order in soft fields,
as the soft fields $W$ and ${\bm A}$ enter in Eq.~(\ref{h}).
\\

[{\it Step 3}]: Substituting these expansions into $\xi_{\pm}$ in Eq.~(\ref{xi}), 
one obtains the expansion
\begin{align} 
\xi_{\pm} &= \xi_{\pm}^{(0)} + \xi_{\pm}^{(1)} + \xi_{\pm}^{(2)} + \cdots,
\end{align}
where, e.g., 
\begin{align}
\label{xi01}
\xi_{+}^{(0)} &= -ig[{\bm v} \cdot {\bm a}^{(0)}, w_{+}^{(0)}]_{\rm soft},
\quad 
\xi_{+}^{(1)} = -ig([{\bm v} \cdot {\bm a}^{(1)}, w_{+}^{(0)}]_{\rm soft}
+ [{\bm v} \cdot {\bm a}^{(0)}, w_{+}^{(1)}]_{\rm soft}), \\
\label{xi0-}
\xi_{-}^{(0)} &= -ig[{\bm v} \cdot {\bm a}^{(0)}, w_{-}^{(0)}]_{\rm soft}
-\frac{ig}{2}\epsilon_{ijk}v^i\partial_t[a^{(0)j},a^{(0)k}]_{\rm soft}, \\
\label{xi1-}
\xi_{-}^{(1)} &= -ig([{\bm v} \cdot {\bm a}^{(1)}, w_{-}^{(0)}]_{\rm soft}
+[{\bm v} \cdot {\bm a}^{(0)}, w_{-}^{(1)}]_{\rm soft})
-ig\epsilon_{ijk}v^i\partial_t[a^{(0)j},a^{(1)k}]_{\rm soft}.
\end{align}
One might expect that $\xi_{\pm} \simeq \xi_{\pm}^{(0)}$ is a good approximation 
in the equations of motion for the soft fields (\ref{W+}) and (\ref{W-}).
However, when one takes the thermal average over initial conditions, one finds
\beq
\langle \xi_{\pm,a}^{(0)} \rangle = g f^{abc} \langle ({\bm v} \cdot {\bm a_b}^{(0)} w_{\pm,c}^{(0)})_{\rm soft} \rangle
\propto f^{abc} \delta_{bc} = 0,
\eeq
where we used $\langle {\bm a_b}^{(0)} w_{\pm,c}^{(0)} \rangle \propto \delta_{bc}$ and 
the antisymmetric property of the structure constant $f^{abc}$. So one also needs to 
take into account the subleading term $\xi^{(1)}_{\pm}$ in the effective theory.
Notice that though $\langle \xi_{\pm}^{(0)} \rangle$ is vanishing,
the two-point function $\langle \xi_{\pm}^{(0)}(x,{\bm v})  \xi_{\pm}^{(0)}(x',{\bm v}') \rangle$
is not necessarily vanishing. 
Actually, $\langle \xi_{+,a}^{(0)}(x,{\bm v})  \xi_{+,b}^{(0)}(x',{\bm v'}) \rangle$
was previously computed in Ref.~\cite{Bodeker:1998hm} to be
\begin{align}
\label{xi-xi}
\langle \xi_{+,a}^{(0)}(x,{\bm v}) \xi_{+,b}^{(0)}(x',{\bm v'}) \rangle = 
\frac{2 T}{3 \sigma_c} I_+({\bm v}, {\bm v}')\delta_{ab}\delta^{(4)}(x-x'),
\end{align}
where
\begin{align}
\label{I}
I_+({\bm v}, {\bm v}') = \delta^{(2)}({\bm v}-{\bm v}')
-\frac{4}{\pi}\frac{({\bm v} \cdot {\bm v}')^2}{\sqrt{1-({\bm v} \cdot {\bm v}')^2}}.
\end{align}
On the other hand, $\langle \xi_{-}^{(0)}(x,{\bm v}) \xi_{-}^{(0)}(x',{\bm v}') \rangle$
will be shown to be irrelevant to our final effective theory, and we will not try to 
compute its detailed form in this paper.

From the expression (\ref{xi01}) and the solutions (\ref{solutions}),
one finds $\langle\xi_+^{(1)}\rangle \sim \# g^2 T W_+ + \# g^2 W_-$ and 
$\langle\xi_-^{(1)}\rangle \sim  \#g^3T^2 W_+ + \#g^2 T W_-$
 [see Eq.~(\ref{xi+-1st}) in Appendix \ref{sec:order_estimates}].
Taking into account the ${\bm v}$ dependence, 
$\xi^{(1)}_{\pm}$ are generally written as
\beq
\label{xi1}
\langle \xi^{(1)}_{\pm}(x,{\bm v})\rangle &=& \delta C_{\pm}[W_{+},W_{-}],\nonumber \\
\delta C_{+}[W_{+},W_{-}] 
&=&  \int_{{\bm v}'} J_{++}({\bm v}, {\bm v}')W_{+}(x,{\bm v}')
+ \int_{{\bm v}'} J_{+-}({\bm v}, {\bm v}')W_{-}(x,{\bm v}'), \nonumber\\
\delta C_{-}[W_{+},W_{-}] 
&=&  \int_{{\bm v}'} J_{-+}({\bm v}, {\bm v}')W_{+}(x,{\bm v}')
+  \int_{{\bm v}'} J_{--}({\bm v}, {\bm v}')W_{-}(x,{\bm v}'),
\eeq
with some functions $J_{\pm\pm}({\bm v}, {\bm v}')$. 
The contribution of $W_+$ in $\delta C_+[W_+,W_-]$ was
previously found to be \cite{Bodeker:1998hm, Arnold:1998cy}
\beq
\gamma  \int_{{\bm v}'} I_{+}({\bm v}, {\bm v}')W_{+}(x,{\bm v}') \equiv \delta C[W_+], 
\eeq
where $\gamma$ is the damping rate given by Eq.~(\ref{gamma}) and
$I_{+}({\bm v}, {\bm v}')$ is given by Eq.~(\ref{I}).
Again, we will see below that $J_{+-}$, $J_{-+}$, and $J_{--}$ are irrelevant to our 
effective theory of interest. Physically, $\delta C_{\pm}$ correspond
to the collision terms and the typical scale of $\delta C_{\pm}$ is the scale of small-angle 
scattering that causes color diffusion, $\gamma \sim g^2 T \ln(1/g)$ \cite{Arnold:1998cy}.
In Appendix \ref{sec:order_estimates}, we evaluate the magnitude of $\xi^{(n)}_{\pm}$ 
following Ref.~\cite{Bodeker:1998hm} and show that an expansion up to $\xi^{(1)}_{\pm}$ 
is enough for our present analysis in the leading order in $g$.

Inserting $\xi_{\pm} \simeq \xi_{\pm}^{(0)} +\langle\xi_{\pm}^{(1)}\rangle$ 
and Eq.~(\ref{xi1}) into the equations of the soft fields, Eqs.~(\ref{W+}) and (\ref{W-}), 
we arrive at
\begin{align}
\label{EOM_W+}
[v \cdot D, W_{+} (x, {\bm v})] &= {\bm v} \cdot {\bm E}(x) 
+ \xi^{(0)}_+(x,{\bm v}) + \delta C_{+}[W_+,W_-], \\
\label{EOM_W-}
[v \cdot D, W_{-} (x, {\bm v})] &= - {\bm v} \cdot \d_t {\bm B}(x) 
+ \xi^{(0)}_-(x,{\bm v}) + \delta C_{-}[W_+,W_-].
\end{align}
\\

[{\it Step 4}]: We can now show that $W_-$ and $\xi_-$ are irrelevant to our 
effective theory at the soft scale to the leading order in $g$.
As we have seen in Sec.~\ref{sec:bottom-up} and as we will show in 
Sec.~\ref{sec:instability} for chiral plasmas, the typical time scale is $\tau \sim (g^4 T)^{-1}$ 
[up to a factor of $\ln(1/g)$]. 
Then $\xi_{+}^{(0)}$ and $\xi_{-}^{(0)}$ can be estimated as
$\xi_{+}^{(0)}(x,{\bm v}) \sim g^5 T^2$ \cite{Bodeker:1998hm} and 
$\xi_{-}^{(0)}(x,{\bm v}) \sim g^6 T^3$, respectively (see Appendix \ref{sec:order_estimates}).
Taking into account $[v \cdot D, W_{\pm}] \sim g^2 T W_{\pm}$ and Eq.~(\ref{xi+-1st})
and noting ${\bm v} \cdot {\bm E} \sim \xi^{(0)}_+$ and
${\bm v} \cdot \d_t {\bm B} \ll \xi^{(0)}_-$,
one can estimate the typical amplitudes of $W_{\pm}$ as
\begin{eqnarray}
\label{W+_order}
W_+(x,{\bm v}) &\sim& \frac{\xi^{(0)}_+}{k} \sim g^3T, \\
\label{W-_order}
W_-(x,{\bm v})  &\sim& \frac{\xi^{(0)}_-}{k} \sim g^4 T^2,
\end{eqnarray}
up to factors of $\ln(1/g)$. 
Then this leads to $\delta C_+[W_+,W_-]\approx \delta C[W_+]$.
In the bracket of Eq.~(\ref{j_W}), the first, second, and third terms are of order
$g^4 T^2$, $g^3 T^2$, and $g^5 T^2$, respectively, and as a result, 
the leading-order contributions to the current read
\begin{align}
\label{j_leading}
{\bm j}^a &= m_D^2\int_{\bm v} {\bm v} W^a_{+}  + 
\frac{N_f \mu_5g^2}{4\pi^2}{\bm B}^a.
\end{align}

We still need to find the form of the first term in Eq.~(\ref{j_leading}), but it was 
already worked out in Refs.~\cite{Bodeker:1998hm, Arnold:1998cy};
in Eq.~(\ref{EOM_W+}), the left-hand side is much smaller than 
$\delta C[W_+]$ by a factor of $\ln(1/g) \gg 1$ and is negligible.%
\footnote{
Precisely speaking, this is not completely justified because
the operator $\delta C$ has a zero mode. One can develop a more rigorous
argument by taking it into account and can check that the result 
remains unchanged in the end \cite{Arnold:1998cy}.
}
The resulting equation is (hereafter $\xi_+^{(0)}$ will be denoted by $\xi$
for simplicity)
\beq
{\bm v} \cdot {\bm E} + \xi \approx \delta C[W_+],
\eeq
and its formal solution is given by
\beq
W_+ = (\delta C)^{-1} ({\bm v} \cdot {\bm E} + \xi),
\eeq
where $\delta C^{-1}$ is understood as an operator that acts on the 
space of functions of ${\bm v}$. 
The first term in Eq.~(\ref{j_leading}) then reduces to \cite{Arnold:1998cy}
\begin{align}
m_D^2\int_{\bm v} {\bm v} W^a_{+} 
= m_D^2 \int_{\bm v} {\bm v} \left[(\delta C)^{-1} ({\bm v} \cdot {\bm E} + \xi)\right]
= \sigma_c {\bm E}^a + {\bm \zeta}^a,
\end{align}
where $\sigma_c = m_D^2/(3\gamma)$ is the color conductivity and 
\beq
{\bm \zeta}^a = 3 \sigma_c \int_{\bm v} {\bm v} \xi
\eeq
is the noise term. Using Eq.~(\ref{xi-xi}), one can show Eq.~(\ref{FD}), which
we asserted on the basis of the fluctuation-dissipation theorem.
\\

[{\it Step 5}]: Finally, to the leading order in $g$, the current 
takes the form
\beq
{\bm j} = \sigma_c {\bm E} + \frac{N_f \mu_5g^2}{4\pi^2}{\bm B} + {\bm \zeta}.
\eeq
Combining it with the spatial part of the Yang-Mills equation (\ref{YM})
(or Amp\`ere's law), one arrives at the chiral Langevin equation
(\ref{chiral_Langevin}) postulated in Sec.~\ref{sec:intuitive}.

\section{Chiral plasma instabilities revisited}
\label{sec:instability}
In light of the chiral Langevin theory we have derived, Eqs.~(\ref{chiral_Langevin}) and (\ref{eq:n_5}), 
we are now ready to study the dynamical evolutions of non-Abelian chiral plasmas. 
First, we notice that the behavior of the mean value of the gauge field in Eq.~(\ref{chiral_Langevin}) 
is governed by the anomalous Yang-Mills equation (the Yang-Mills equation plus non-Abelian 
analogue of the chiral magnetic effect). From the argument in Sec.~\ref{sec:physics}, 
one finds that it exhibits a chiral plasma instability and gauge fields grow rapidly.

We can also estimate the typical time scale of the chiral plasma instability.
From Eq.~(\ref{chiral_Langevin}), it is easy to see that
\beq
\label{tau_inst}
\tau_{\rm inst} \sim \frac{\sigma_c}{k^2} \sim \frac{1}{g^4 T \ln(1/g)}.
\eeq
This is the same time scale as Eq.~(\ref{tau}) without anomalous effects, because in 
the effective theory there is only one length scale $R \sim k_{\rm soft}^{-1} \sim (g^2 T)^{-1}$ 
and the scale of color diffusion $\sigma_c$ for $\mu_5 \sim T$. This result is also 
consistent with the time scale previously obtained in Ref.~\cite{Akamatsu:2013pjd} 
based on the Boltzmann-Vlasov equation with Berry curvature corrections, where the
importance of the color diffusion was found (see footnote 7 in this paper).
Note that the analysis of Ref.~\cite{Akamatsu:2013pjd} is based on the linear response 
theory and is justified only when the nonlinear gauge interactions can be ignored, 
while the present result using the chiral Langevin theory is general in that it is valid 
even when the nonlinear interactions are comparable to the linear ones.

This time scale in turn provides the typical scale of the amplitude of the color 
electric field as in Eq.~(\ref{E}): $E \sim g^5 T^2 \ln(1/g)$. 
Combining it with the amplitude of the color magnetic field,
$B \sim g^3 T^2$, and from Eq.~(\ref{eq:n_5}), 
we can estimate the typical time scale at which the chiral charge $N_5$ varies:
\beq
\tau_{N_5} \sim \frac{T^3}{g^2 EB} \sim \frac{1}{g^{10} T \ln(1/g)}.
\eeq
This is much larger than $\tau_{\rm inst}$, and hence, $\mu_5$ almost ``freezes'' 
during the time $\tau_{\rm inst}$, at least within the applicability of this effective theory. 
Therefore, it justifies the very first assumption that the chiral charge $N_5$ can 
be regarded as conserved during the typical time scale of the system 
and that the chiral chemical potential $\mu_5$ is well defined.

There is an alternative way to see that  $\mu_5$ is well defined in the regime under 
consideration. Rewrite Eq.~(\ref{anomaly}) as
\beq
\label{anomaly2}
\d_t (N_5 + N_{\rm CS}) = 0, 
\eeq
where
\beq
N_5 \equiv \int d^3{\bm x}\, j^{05},
\qquad N_{\rm CS} \equiv \int d^3{\bm x}\, n_{\rm CS},
\eeq
are the global chiral charge and Chern-Simons number, respectively. Equation (\ref{anomaly2}) 
suggests that the combination $N_5 + N_{\rm CS}$ is a conserved charge, so one can 
safely introduce a chemical potential associated with it, which we denote as $\mu_5'$. 
Then we consider the saturation regime where the distribution of the gauge field is the 
equilibrium distribution with fixed $T$ and $\mu'_5$,
\beq
P_{\rm eq}({\bm A})&\sim& e^{-H_{\rm eff}(\bm A)/T}.
\eeq
Here $H_{\rm eff}$ is given in Eq.~(\ref{CS}) with the replacement $\mu_5 \rightarrow \mu_5'$.
From the condition of the equilibrium distribution, $H_{\rm eff}|_{k \sim g^2T} \sim T$
for $\mu'_5\sim T$, one can easily estimate the amplitude of the gauge field as $A \sim gT$.
(Here we assumed that saturation occurs at the magnetic scale, $k \sim g^2 T$.)
The magnitude of the Chern-Simons number is then estimated as 
\beq
N_{\rm CS} \sim g^2 k A^2 \sim g^6 T^3.
\eeq
On the other hand, the chiral charge carried by fermions is $N_5 \sim T^3$; 
only an $O(g^6)$ fraction of the conserved charge $N'_5$ is carried by the gauge field.
So, even if the system starts from a given initial condition, the saturation regime is 
characterized by $\mu_5=[1 + O(g^6)]\mu_5' \sim T$, 
indicating that $\mu_5$ is well defined.

It should be remarked that the separations of scales, $\tau_{\rm inst} \ll \tau_{N_5}$
and $N_{\rm CS} \ll N_5$, we have shown for $g \ll 1$, do not necessarily exist for 
$g \sim 1$; if not, one may \emph{not} define $\mu_5$ itself. 
Indeed, in the quark-gluon plasma created in real heavy ion collision experiments, 
the QCD coupling constant is no longer small, $g \sim 1$, and it is not clear 
at all whether one can define $\mu_5$.  A similar discussion in a different context
is given in Ref.~\cite{Moore:2010jd}.

Note also that our chiral Langevin theory is applicable all the way to saturation as long 
as the definite separation of scales characterized by the small coupling constant $g \ll 1$ exists.
On the other hand, the presence of the chiral plasma instability means that the 
prefactor $\lambda$ defined by $A = \lambda gT$ grows very rapidly. 
In our paper, we have always assumed that $\lambda \sim 1$ 
(which could be a large factor not captured by the expansion in $g$) 
and that $\lambda g \ll 1$. Beyond the counting scheme of the present paper,
one could imagine the situation that the large $\lambda$ overwhelms the small $g$ such 
that $\lambda g \gtrsim 1$. It would be interesting to consider a possible effective theory 
description in this regime where the naive expansion in $g$ breaks down. Such an effective 
theory might give new insight into the physics of the chiral plasma instability towards the 
saturation regime.

\section{Conclusion}
\label{sec:conclusion}
In this paper, we have constructed a Langevin-type effective theory that describes the dynamics 
of non-Abelian plasmas with chirality imbalance at the magnetic scale $\sim g^2 T$. 
This chiral Langevin theory, in particular, describes the evolution of the chiral plasma instability
towards saturation, which is completely missed in hydrodynamics for non-Abelian plasmas.
Using this equation, the time scale of the chiral plasma instability is easily found to be
$\tau_{\rm inst} \sim [g^4 T \ln(1/g)]^{-1}$, as is consistent with the estimate previously 
found in Ref.~\cite{Akamatsu:2013pjd} based on the Boltzmann-Vlasov equation with 
Berry curvature corrections. On the other hand, the time scale of the variation of the chiral charge 
is $\tau_{N_5} \sim [g^{10} T \ln(1/g)]^{-1}$ and is much longer than $\tau_{\rm inst}$; 
the chiral charge is thus shown to be approximately conserved during the chiral plasma instability, 
which ensures that $\mu_5$ is well defined for $g \ll 1$.

In this paper, we have derived the chiral Langevin equation both from a physical argument 
and from a microscopic analysis. For the latter, we started with the linearized non-Abelian 
Boltzmann-Vlasov equation with Berry curvature corrections and integrated out (semi)hard 
degrees of freedom. Alternatively, one should also be able to arrive at the same Langevin-type 
equation starting from a different classical transport theory \cite{Kelly:1994ig} with Berry 
curvature corrections \cite{Stone:2013sga} where the trajectories of a particle are specified 
by ${\bm x}$, ${\bm p}$, and the non-Abelian charge $Q^a$ (known as the Wong equations \cite{Wong:1970fu}). 
Such a derivation was done in the case without anomalous parity-violating effects 
in Ref.~\cite{Litim:1999ns}.

A detailed numerical analysis of the chiral Langevin theory should allow us to understand the 
numerical coefficient in Eq.~(\ref{tau_inst}), how the system approaches saturation after the 
chiral plasma instability, and what the configuration at the saturation stage looks like. 
These investigations are deferred to future work \cite{Akamatsu}.

\acknowledgments
The authors thank D.~B\"odeker, C.~Manuel, S.~Pu, K.~Rajagopal, A.~Rothkopf, and 
D.~T.~Son for discussions. One of the authors (N.Y.) also thanks the hospitality of 
IFT UAM-CSIC, where a part of this work was carried out.

\appendix

\section{Fokker-Planck equation}
\label{sec:FP}
For completeness, we review the derivation of the Fokker-Planck equation from the 
Langevin equation (\ref{Langevin}). The Fokker-Planck equation here describes how 
the probability function for the gauge field ${\bm A(t,\bm x)}$ relaxes to the thermal 
equilibrium distribution $e^{-H_{\rm eff}/T}$ over a long time span. 
Our derivation essentially follows Ref.~\cite{Chaikin}.

We first introduce the probability
\beq
\label{P}
P\left[{\bm A(\bm x)},  t|{\bm A}_0(\bm x), t_0\right] = 
\langle \delta\left[{\bm A}(\bm x)-{\bm A}(t,\bm x)\right] \rangle_{{\bm A}_0,t_0},
\eeq
that the gauge field has the configuration ${\bm A}(\bm x)$ at time $t$, given the initial 
gauge field configuration ${\bm A}_0(\bm x)$ at time $t_0$. Then, the probability function satisfies
\beq
P\left[{\bm A}(\bm x),  t + \Delta t|{\bm A}_0(\bm x), t_0\right] = 
\int {\mathcal D}{\bm A}' P\left[{\bm A}(\bm x),  t + 
\Delta t|{\bm A}'(\bm x), t\right] P\left[{\bm A}'(\bm x), t|{\bm A}_0(\bm x), t_0\right],
\eeq
with infinitesimally small $\Delta t$, where
\beq
\label{P_cond}
P\left[{\bm A}(\bm x),  t + \Delta t|{\bm A}'(\bm x), t\right] = 
\langle \delta\left[{\bm A}(\bm x)-{\bm A}(t + \Delta t,\bm x)\right] \rangle_{{\bm A}',t}.
\eeq
Below we shall compute it explicitly.

Using the Langevin equation~(\ref{Langevin}), we have
\beq
\label{integral}
{\bm A}(t + \Delta t,\bm x) = {\bm A}'(\bm x) - 
\frac{\Delta t}{\sigma}\frac{\delta H_{\rm eff}(\bm A')}{\delta {\bm A}'(\bm x)} 
+ \frac{1}{\sigma}\int_{t}^{t + \Delta t} dt' {\bm \zeta}(t',\bm x).
\eeq
Then we substitute it into Eq.~(\ref{P_cond}) and expand up to linear terms in $\Delta t$. 
In this process, note that the average of the third term on the right-hand side
of Eq.~(\ref{integral}) is vanishing, while the average of its square is
\beq
\int_{t}^{t + \Delta t} dt_1 \int_{t}^{t + \Delta t} dt_2
\langle \zeta_i(t_1,\bm x_1) \zeta_j(t_2,\bm x_2) \rangle = 2\sigma T \delta_{ij} \Delta t\delta(\bm x_1 - \bm x_2).
\eeq
Note also that averages of higher-order terms in $\int dt \zeta_i(t,{\bm x})$ are higher order 
in $\Delta t$ and are negligible. This is because they can be expressed as products of 
$\langle \zeta_i(t_1,\bm x_1) \zeta_j(t_2,\bm x_2) \rangle \sim \delta_{ij}\Delta t^{-1}\delta(\bm x_1-\bm x_2)$ 
due to the fact that ${\bm \zeta}$ is a Gaussian white noise. We thus have
\beq
&&\langle \delta\left[{\bm A}-{\bm A}(t + \Delta t)\right] \rangle_{{\bm A}',t}\nonumber\\
&&= \left[1 +\Delta t \frac{T}{\sigma}
\int d^3{\bm x}\, \left(\frac{1}{T} \frac{\delta H_{\rm eff}(\bm A')}{\delta {\bm A}'}
\cdot \frac{\delta}{\delta {\bm A}} 
+ \frac{\delta^2}{\delta {\bm A^2}}\right) \right]
\delta\left[{\bm A}-{\bm A}'\right],
\eeq
where the argument ${\bm x}$ of ${\bm A}$ is suppressed for notational simplicity.

Remembering the definition of $P$ in Eq.~(\ref{P}), we arrive at
\beq
\frac{\d P}{\d t} =\frac{T}{\sigma}
\int d^3{\bm x}\, \frac{\delta}{\delta {\bm A}} \cdot
\left[
\left(\frac{1}{T} \frac{\delta H_{\rm eff}(\bm A)}{\delta {\bm A}}
+ \frac{\delta}{\delta {\bm A}}
\right) P\right]
\eeq
This is the Fokker-Planck equation.
From the condition $\partial P/\partial t = 0$, we obtain the equilibrium distribution
\beq
P_{\rm eq} \sim e^{-H_{\rm eff}/T}.
\eeq
We note that the derivation here does not depend on the details of $H_{\rm eff}$
and is valid for both $H_{\rm eff}$'s given in Eqs.~(\ref{MS}) and (\ref{CS}).

\section{Solution to the equations of motion for semihard modes}
\label{sec:solution}
In this appendix, we give a detailed procedure to solve 
Eqs.~(\ref{EOM_w+}), (\ref{EOM_w-}), and (\ref{EOM_aT}) 
with the use of the one-sided Fourier transform.
The one-sided Fourier transform is defined for a function $f(t)$ by
\beq
f(k_0) = \int_0^{\infty} dt e^{i k_0 t} f(t).
\eeq
From the definition, it is easy to derive the relations
\begin{align}
\int_0^{\infty} dt e^{i k_0 t} \d_t f(t) &= -f_{\rm in} -ik_0 f(k_0), \\
\int_0^{\infty} dt e^{i k_0 t} \d_t^2 f(t) &= - \d_t f_{\rm in} + ik_0 f_{\rm in} -k_0^2 f(k_0),
\end{align}
where the subscript ``in" stands for the initial values at $t=0$.
Using these relations for Eqs.~(\ref{EOM_w+}) and (\ref{EOM_w-}), we can express $w_{\pm}$
in terms of ${\bm a}$ as
\begin{align}
\label{w+}
w_+(K,{\bm v}) &= \frac{1}{v \cdot K}\left[-k_0 {\bm v} \cdot {\bm a}(K) +il_+(K,{\bm v})\right], \\
\label{w-}
w_-(K,{\bm v}) &= \frac{1}{v \cdot K}\left[-ik_0{\bm v} \cdot ({\bm k} \times {\bm a}(K)) +il_-(K,{\bm v})\right],
\end{align}
where $K^{\mu}=(k_0,{\bm k})$ and 
\begin{align}
\label{l+}
l_+(K,{\bm v}) & \equiv w_+^{\rm in} ({\bm k}, {\bm v}) + {\bm v} \cdot {\bm a}^{\rm in}({\bm k})
 + h_+(K, {\bm v}), \\
\label{l-}
l_-(K,{\bm v}) & \equiv w_-^{\rm in} ({\bm k}, {\bm v}) + i{\bm v} \cdot ({\bm k} \times {\bm a}^{\rm in}({\bm k}))
+ h_-(K, {\bm v}).
\end{align} 
Substituting these expressions into the one-sided Fourier transform of Eq.~(\ref{EOM_aT}), 
one has
\begin{align}
\label{a}
- K^2 {\bm a}_T(K) - m_D^2\int_{\bm v} {\bm v}_T w_{+}(K, {\bm v}) 
- \frac{N_f g^2 \mu_5}{4\pi^2}\int_{\bm v} {\bm v}_T w_{-}(K, {\bm v}) = {\bm l}(K),
\end{align}
with
\begin{align}
\label{l}
{\bm l}(K) & \equiv -{\bm e}_{T}^{\rm in}({\bm k}) - ik_0 {\bm a}_{T}^{\rm in}({\bm k}).
\end{align}
In Eq.~(\ref{a}), the contribution from the first term in Eq.~(\ref{w-}) is negligible
compared with the contributions from $w_+$.
One can then solve this equation in terms of ${\bm a}$, and as a result, 
one can write $w_{\pm}$ as
\begin{subequations}
\label{solutions_l}
\begin{align}
a^i(K) &= \Delta^{ij}_{11}(K)l^j(K) + \int_{\bm v} \Delta^i_{12}(K,{\bm v}) 
\left[l_+(K,{\bm v})  + \frac{N_f g^2 \mu_5}{4\pi^2 m_D^2} l_-(K,{\bm v}) \right], \\
w_+(K,{\bm v}) &= \Delta^i_{21}(K,{\bm v})l^i(K) 
+  \int_{\bm v'} \left[ \Delta_{22}(K,{\bm v},{\bm v}') l_+(K,{\bm v'}) 
+ \frac{N_f g^2 \mu_5}{4\pi^2 m_D^2} \Delta_{23}(K,{\bm v},{\bm v}') l_-(K,{\bm v'}) \right], \\
w_-(K,{\bm v}) &= \Delta_{31}^{i}(K,\bm v)l^i(K)
+\int_{\bm v'} \Delta_{32}(K,{\bm v},{\bm v}') l_+(K,{\bm v'}) +\Delta_{33}(K,{\bm v})l_-(K,{\bm v}),
\end{align}
\end{subequations}
where the propagators $\Delta^{ij}_{11}$, $\Delta^{i}_{21}$, and $\Delta^{i}_{31}$ are defined as
\begin{subequations}
\begin{align}
\Delta^{ij}_{11}(K) &= P^{ij}_T \Delta_T(K), \\
\Delta^{i}_{21}(K,{\bm v}) & = -\frac{k_0}{v \cdot K}\Delta_T(K) v_T^i, \\
\Delta^{i}_{31}(K,{\bm v}) & = \frac{ik_0}{v \cdot K}\Delta_T(K) \epsilon^{ijk} k^j v_T^k, 
\end{align}
\end{subequations}
and $\Delta^{i}_{12}$, $\Delta_{22}$, $\Delta_{23}$, $\Delta_{32}$, and $\Delta_{33}$ 
are defined in Eq.~(\ref{propagators}).
Substituting the expressions for $l_{\pm}$ and ${\bm l}$ into 
Eq.~(\ref{solutions_l}), one obtains Eq.~(\ref{solutions}).

\section{Magnitudes of $\xi_{\pm}^{(n)}$}
\label{sec:order_estimates}

In this appendix, we briefly sketch the estimation of $\xi_{\pm}^{(n)}$ following the 
method in Ref.~\cite{Bodeker:1998hm}. We will perform the analysis in the $K$ space.

First let us start from the initial amplitudes of the semihard modes: 
\beq
\label{a_in}
{a}^{\rm in}({\bm k})\sim w^{\rm in}_{+}({\bm k},{\bm v})\sim g^{-5/2}T^{-2}, \quad
w^{\rm in}_{-}({\bm k},{\bm v})\sim g^{-3/2}T^{-1}
\eeq
for $k\sim gT$.
The former derives from the thermal average with the hard thermal loop effective Hamiltonian 
\cite{Bodeker:1998hm} and the latter from Eq.~(\ref{BV_w-}), $w_{-}(x,{\bm v})\sim \partial a(x)$.
Then the semihard modes at the zeroth order in soft fields can be estimated,
using Eq.~(\ref{solutions_l}) with $h_{\pm}=0$, 
as ${a}^{(0)}(K)\sim w^{(0)}_{+}(K,{\bm v})\sim g^{-7/2}T^{-3}$ and 
$w^{(0)}_{-}(K,{\bm v})\sim g^{-5/2}T^{-2}$.
The statistical averages of these modes are
\begin{subequations}
\label{average}
\begin{align}
\langle a^{(0)}(K)a^{(0)}(K')\rangle
&\sim\langle w^{(0)}_{+}(K,{\bm v})a^{(0)}(K')\rangle\sim\langle 
w^{(0)}_{+}(K,{\bm v})w^{(0)}_{+}(K',{\bm v'})\rangle\nonumber \\
&\sim g^{-7}T^{-6}\sim g^{-3}T^{-2}\delta^{(4)}(K+K'),\\
\langle w^{(0)}_{-}(K,{\bm v})a^{(0)}(K')\rangle
&\sim\langle w^{(0)}_{-}(K,{\bm v})w^{(0)}_{+}(K',{\bm v'})\rangle\nonumber \\
&\sim g^{-6}T^{-5}\sim g^{-2}T^{-1}\delta^{(4)}(K+K'),\\
\langle w^{(0)}_{-}(K,{\bm v})w^{(0)}_{-}(K',{\bm v'})\rangle
&\sim g^{-5}T^{-4}\sim g^{-1}\delta^{(4)}(K+K'),
\end{align}
\end{subequations}
where the $\delta$-functions are used to ensure the energy-momentum conservation.%
\footnote{
To be precise, the $\delta$ function for the frequency $k^0$ should be understood 
as ${i}/{(2\pi k^0)}$ because we perform the one-sided Fourier transformation.
The same remark applies below.
}

Second let us observe the property of factorization \cite{Bodeker:1998hm}.
In general, the semihard contribution to the soft sector of $\xi(P)$ ($\xi=\xi_{\pm}$) 
can be written as [see Eq.~(\ref{xi})]
\begin{eqnarray}
\xi(P)\sim g\int\frac{d^4 K}{(2\pi)^4}\phi(K)\phi(P-K),
\end{eqnarray}
with $\phi = a, w_{\pm}$, where $p\sim g^2T$ is soft (with $p^0$ undetermined)
while $K\sim gT$ is semihard.
Then the solution $\phi=\phi^{(0)}+\phi^{(1)}+\cdots$ provides the expansion of $\xi$ 
in terms of soft fields, $\xi=\xi^{(0)}+\xi^{(1)}+\cdots$.
Note that $\phi^{(n+1)}$ is iteratively obtained by convoluting $\phi^{(n)}$ ($n=0,1,2,\cdots$)
with the soft fields $\Phi=A,W_{\pm}$, and is linear in $\phi^{(n)}$,
\begin{eqnarray}
\phi^{(n+1)}(K)\sim c_{\Phi}(K)\int\frac{d^4 P'}{(2\pi)^4}
\phi^{(n)}(K-P')\Phi(P'),
\end{eqnarray}
with some coefficient $c_{\Phi}(K)$. 
By introducing a new variable $\chi\equiv\phi^{(0)}(K-P')\phi^{(0)}(P-K)$, 
where the sum of the arguments of $\phi^{(0)}$ is soft, $N$-point functions of $\xi$ are 
expressed in the form of $\langle\chi_1\chi_2\cdots\chi_N\rangle$.
In this $N$-point function, each disconnected part yields a $\delta$ function 
for soft momentum, $\delta^{(4)}(P)\sim (p^0)^{-1}(g^2T)^{-3}$, instead of that for 
semihard momentum, $\delta^{(4)}(K)\sim (gT)^{-4}$.
Thus we obtain 
\begin{eqnarray}
\langle\chi_1\chi_2\cdots\chi_N\rangle_{\rm conn}
&\sim& g^2 \frac{p^0}{T}\langle\chi_1\cdots\chi_M\rangle_{\rm conn}
\langle\chi_{M+1}\cdots\chi_N\rangle_{\rm conn}\nonumber \\
&\ll&\langle\chi_1\cdots\chi_M\rangle_{\rm conn}\langle\chi_{M+1}\cdots\chi_N\rangle_{\rm conn}
\ll \langle\chi_1\rangle\cdots\langle\chi_N\rangle.
\end{eqnarray}
Therefore the approximation $\xi^{(n)}\simeq\langle\xi^{(n)}\rangle$ is enough to 
calculate correlation functions of $\xi$. As mentioned in the main text, 
the exception is $\xi^{(0)}$ because $\langle\xi^{(0)}\rangle=0$, which
we need to introduce as a Gaussian noise.%
\footnote{
One can show that the Gaussian noise $\xi^{(0)}$ gives a dominant 
contribution by noting, e.g., $\left[\langle\xi^{(0)}\xi^{(0)}\rangle_{\rm conn}\right]^{1/2}
\gg\left[\langle\xi^{(0)}\xi^{(0)}\xi^{(0)}\rangle_{\rm conn}\right]^{1/3}$ \cite{Bodeker:1998hm}.
}

We now evaluate the magnitudes of $\xi^{(n)}_{\pm}(P,{\bm v})$. 
The time scale of our interest will be $\tau \sim (g^4 T)^{-1}$ [see Eq.~(\ref{tau_inst})],
and so we consider the regime with $p^0\sim g^4T$ and $p\sim g^2T$.
The parity-even parts,
$\xi_{+}^{(0)}$ and $\langle\xi_{+}^{(1)}\rangle$, were already evaluated
in Ref.~\cite{Bodeker:1998hm}, 
which we will rederive below. Here we shall also estimate the parity-odd ones,
$\langle \xi_{\pm}^{(0)}(P,{\bm v}) \rangle$ and $\xi_{\pm}^{(0)}(P,{\bm v})$.
By putting $K' = P-K$ in Eq.~(\ref{average}), we have
\begin{subequations}
\begin{eqnarray}
\langle{a^{(0)}(K)}w_{+}^{(0)}(P-K)\rangle &\sim 
g^{-3}T^{-2}\delta^{(4)}(P) \sim g^{-13} T^{-6},
\\
\langle{a^{(0)}(K)}w_{-}^{(0)}(P-K)\rangle &\sim 
g^{-2}T^{-1}\delta^{(4)}(P) \sim g^{-12} T^{-5},
\\
\langle{a}^{(0)}(K){a}^{(0)}(P-K)\rangle &\sim g^{-3}T^{-2}\delta^{(4)}(P)
\sim g^{-13} T^{-6},
\end{eqnarray}
\end{subequations}
from which the nonlinear term of $a$ in the statistical average of Eq.~(\ref{xi0-}) is 
found to be negligible compared with the linear one. We thus have
\begin{subequations}
\begin{eqnarray}
\langle\xi_{+}^{(0)}(P,{\bm v})\rangle
&\sim& g\int\frac{d^4 K}{(2\pi)^4}
\langle{a^{(0)}(K)}w_{+}^{(0)}(P-K)\rangle \sim g^{-8} T^{-2},\\
\xi_{+}^{(0)}(P,{\bm v})&\sim& [\langle \xi_{+}^{(0)}(P,{\bm v})\xi_{+}^{(0)}(P',{\bm v'})\rangle_{\rm conn}]^{1/2}
\sim \left(g^2 \frac{p^0}{T} \right)^{\! \! 1/2} \! \! \langle\xi_{+}^{(0)}(P,{\bm v})\rangle \sim g^{-5} T^{-2},\\
\langle\xi_{-}^{(0)}(P,{\bm v})\rangle
&\sim& g\int\frac{d^4 K}{(2\pi)^4}
\langle{a^{(0)}(K)}w_{-}^{(0)}(P-K)\rangle \sim g^{-7} T^{-1},\\
\xi_{-}^{(0)}(P,{\bm v})&\sim& [\langle \xi_{-}^{(0)}(P,{\bm v})\xi_{-}^{(0)}(P',{\bm v'})\rangle_{\rm conn}]^{1/2}
\sim \left(g^2 \frac{p^0}{T} \right)^{\! \! 1/2} \! \! \langle\xi_{-}^{(0)}(P,{\bm v})\rangle \sim g^{-4} T^{-1}. \ 
\end{eqnarray}
\end{subequations}
In position space, they correspond to 
\begin{subequations}
\begin{eqnarray}
\label{xi+}
\langle \xi_{+}^{(0)}(x,{\bm v}) \rangle \sim g^2 T^2, \qquad 
\xi_{+}^{(0)}(x,{\bm v}) \sim g^5 T^2, \\
\label{xi-}
\langle \xi_{-}^{(0)}(x,{\bm v}) \rangle \sim g^3 T^3, \qquad 
\xi_{-}^{(0)}(x,{\bm v}) \sim g^6 T^3.
\end{eqnarray}
\end{subequations}

We then turn to the magnitudes of $\xi_{\pm}^{(1)}(P,\bm v)$ given in Eqs.~\eqref{xi01} and \eqref{xi1-}.
Using Eq.~\eqref{average}, the contribution of $W_+(P,\bm v)$ in 
$\langle\xi_{+}^{(1)}(P,\bm v)\rangle$ is found to be
\begin{eqnarray}
\label{xi+1stW+}
&&\frac{g}{T}\int\frac{d^4K}{(2\pi)^4}\frac{d^4K'}{(2\pi)^4}
\left[\# \langle a^{(0)}(K')a^{(0)}(P-K) \rangle + \# \langle a^{(0)}(K')w_+^{(0)}(P-K,\bm v) \rangle
\right]
{W_+(K-K',\bm v)}\nonumber \\
&&\sim
\frac{g}{T} (gT)^8 (g^{-7}T^{-6}) W_+(P,\bm v) \sim g^2TW_+(P,\bm v).
\end{eqnarray}
Here and below ``$\#$'' stands for some coefficients of order $O(g^0)$.
We can obtain the other contributions similarly, and 
eventually arrive at
\begin{subequations}
\label{xi+-1st}
\begin{align}
\langle\xi_{+}^{(1)}(P,\bm v)\rangle &\sim
\#g^2TW_+(P,\bm v) + \#g^2W_-(P,\bm v),\\
\langle\xi_{-}^{(1)}(P,\bm v)\rangle &\sim
\#g^3T^2W_+(P,\bm v) + \#g^2TW_-(P,\bm v).
\end{align}
\end{subequations}
By using the estimates of $W_+$ and $W_{-}$ in Eqs.~(\ref{W+_order}) and (\ref{W-_order})
[which are obtained based on Eqs.~(\ref{EOM_W+}) and (\ref{EOM_W-})], we have
\begin{subequations}
\begin{eqnarray}
\langle\xi_{+}^{(1)}(P,{\bm v})\rangle
\sim \xi_{+}^{(0)}(P,{\bm v})\sim g^{-5}T^{-2}, \\
\langle\xi_{-}^{(1)}(P,{\bm v})\rangle
\sim \xi_{-}^{(0)}(P,{\bm v})\sim g^{-4}T^{-1}.
\end{eqnarray}
\end{subequations}

\begin{table}[t]
\caption{The frequency spectra of the soft fields in the $P$ space 
(with $p \sim g^2 T$).} 
\begin{tabular}{|c|c|c|c|}
\hline 
$p_0$   & $A(P)$ & $W_+(P,{\bm v})$ & $W_-(P,{\bm v})$  \\ \hline \hline
$g^4T$  & $g^{-9}T^{-3}$ & $g^{-7}T^{-3}$ & $g^{-6}T^{-2}$     \\
$g^2T$  & $g^{-6}T^{-3}$ & $g^{-6}T^{-3}$ & $g^{-5}T^{-2}$      \\
$gT$      & $g^{-5}T^{-3}$ & $g^{-5}T^{-3}$ & $g^{-4}T^{-2}$     \\
\hline
\end{tabular}
\label{table:spectra}
\end{table}

By repeating this procedure, one can also evaluate the higher-order terms, 
$\langle \xi^{(n\geq2)}_{\pm} (P,{\bm v}) \rangle$. For this purpose, one needs 
to know the ``frequency spectra'' of the soft fields $W_{\pm}(P,\bm v)$. 
The amplitudes of the soft fields for $p_0 \lesssim g^2 T$ can be obtained similarly 
to the argument leading to Eq.~(\ref{a_in}), while those for $p_0 \sim g^4T$ can be 
obtained by using the equations of motion for $W_{\pm}$, 
Eqs.~(\ref{EOM_W+}) and (\ref{EOM_W-}).
The amplitudes of $A(P)$ and $W_{+}(P,\bm v)$ (already obtained in Ref.~\cite{Bodeker:1998hm}) 
together with that of $W_{-}(P,\bm v)$ are summarized in Table \ref{table:spectra}. 
From this, one can estimate the upper bounds of 
$\langle\xi_{\pm}^{(n\geq2)}(P,{\bm v})\rangle$ as follows:
\begin{subequations}
\begin{eqnarray}
\langle\xi_{+}^{(n\geq 2)}(P,\bm v)\rangle &\sim& 
g^n (gT)^{2-n} \int_{P_1} \cdots \int_{P_{n-1}} 
W_+(P_1,\bm v) \cdots W_+(P_{n-1},\bm v) W_+(P',\bm v)
\nonumber \\
&\lesssim& g^n (gT)^{2-n} (g^2 T)^{n-1} (g^{-7} T^{-3}) \sim g^{2n-7}T^{-2}, \\
\langle\xi_{-}^{(n\geq 2)}(P,\bm v)\rangle &\sim& 
g^n (gT)^{3-n} \int_{P_1} \cdots \int_{P_{n-1}} 
W_+(P_1,\bm v) \cdots W_+(P_{n-1},\bm v) W_+(P',\bm v)
\nonumber \\
&\lesssim& g^n (gT)^{3-n} (g^2 T)^{n-1} (g^{-7} T^{-3})
\sim g^{2n-6}T^{-1},
\end{eqnarray} 
\end{subequations}
where $\int_{P} \equiv \int d^4 P$ and $P' \equiv P-P_1-\cdots-P_{n-1}$, 
and we used $\int_{P_k} W_+(P_k) \sim g^2 T$ ($k=1,\cdots,n-1$) and
$W_+(P') \lesssim g^{-7} T^{-3}$.
Therefore, $\langle\xi_{\pm}^{(n\geq2)}(P,{\bm v})\rangle$ are much smaller 
than the zeroth- and first-order terms for $g \ll 1$, and can indeed be ignored in 
Eqs.~\eqref{EOM_W+} and \eqref{EOM_W-} as we postulated.

\end{document}